\documentclass[lettersize,journal]{IEEEtran}
\usepackage{amsmath,amsfonts}
\usepackage{algorithmic}
\usepackage{algorithm}
\usepackage{array}
\usepackage[caption=false,font=normalsize,labelfont=sf,textfont=sf]{subfig}
\usepackage{textcomp}
\usepackage{stfloats}
\usepackage{url}
\usepackage{verbatim}
\usepackage{graphicx}
\usepackage{cite}
\usepackage{microtype}
\usepackage{graphicx}
\usepackage{booktabs} % for professional tables
\usepackage{bbm}
\usepackage{hyperref}
\usepackage{algorithm}
\usepackage[algo2e]{algorithm2e}
\usepackage{amsfonts}
\usepackage{newfloat}
\usepackage{listings}
\usepackage{stmaryrd}
\usepackage{multirow}
\usepackage{amsmath}
\usepackage{amssymb}
\usepackage{mathtools}
\usepackage{amsthm}
\usepackage{bbding}
\usepackage{pifont}
\usepackage{wasysym}
\usepackage{threeparttable}
\usepackage[capitalize,noabbrev]{cleveref}
\theoremstyle{plain}
\newtheorem{theorem}{Theorem}[section]

\theoremstyle{definition}
\newtheorem{definition}[theorem]{Definition}

\theoremstyle{remark}

\newenvironment{packeditemize}{
\begin{list}{$\bullet$}{
\setlength{\labelwidth}{8pt}
\setlength{\itemsep}{0pt}
\setlength{\leftmargin}{\labelwidth}
\addtolength{\leftmargin}{\labelsep}
\setlength{\parindent}{0pt}
\setlength{\listparindent}{\parindent}
\setlength{\parsep}{0pt}
\setlength{\topsep}{3pt}}}{\end{list}}
\newcommand{\sysname}{\textit{OnePath}\xspace}
\usepackage{tikz}
\usepackage{caption}
\usepackage{makecell}
\begin{document}
\title{\sysname: Efficient and Privacy-Preserving Decision Tree Inference in the Cloud\\
% {\color{red}\sysname: A Secure and Efficient Framework for Decision Tree Inference in the Cloud}
}
\author{Shuai Yuan, Rui Zhang, Hongwei Li, ~\IEEEmembership{Fellow,~IEEE}, \\ Xinyuan Qian (Corresponding Author), Guowen Xu, ~\IEEEmembership{Senior Member,~IEEE}

\IEEEcompsocitemizethanks{\IEEEcompsocthanksitem Shuai~Yuan is with the school of Resources and Environment, University of Electronic Science and Technology of China, Chengdu 611731, China.(e-mail: mk2456mk@gmail.com) 
\IEEEcompsocthanksitem Rui~Zhang, Hongwei~Li, Xinyuan~Qian, and Guowen~Xu are with the school of Computer Science and Engineering,  University of Electronic Science and Technology of China, Chengdu 611731, China.(ruizhangsec@163.com; hongweili@uestc.edu.cn; xinyuanqian@outlook.com; guowen.xu@uestc.edu.cn)
}}

\maketitle

\begin{abstract}
The vast storage capacity and computational power of cloud servers have led to the widespread outsourcing of machine learning inference services. While offering significant operational benefits, this practice also introduces privacy risks, such as the exposure of proprietary models and sensitive user data.
In this paper, we present \sysname, a framework for secure and efficient decision tree inference in cloud environments. Unlike existing methods that traverse all internal nodes of a decision tree, our traversal protocol processes only the nodes on the prediction path, significantly improving inference efficiency while preserving privacy. To further optimize privacy and performance, \sysname is the first to employ functional encryption for evaluating decision tree nodes. Notably, our protocol enables both model providers and users to remain offline during the inference phase, offering a crucial advantage for practical deployment. We provide formal security analysis to demonstrate that \sysname provides comprehensive privacy protections during the model inference process. Extensive experimental results show that our approach processes query data in microseconds, highlighting its efficiency. \sysname offers a practical solution that strikes a balance between security and performance, making it a promising option for a wide range of cloud-based decision tree inference applications.
\looseness=-1
\end{abstract}


\section{Introduction}
Machine learning has become a crucial tool for automated data analysis in various domains, including face recognition \cite{sharma2020face} \cite{tabassum2022human}, medical diagnosis \cite{richens2020improving} \cite{swanson2023patterns}, and spam detection \cite{makkar2020efficient} \cite{bhardwaj2023email}. 
Among various machine learning models, decision trees (DT) stand out in classification and regression tasks due to their efficient inference speed, intuitive interpretability, and ability to handle non-linear data. Consequently, they have become one of the most popular models and are widely applied in fields such as intrusion detection \cite{rai2016decision} and medical diagnosis \cite{ghiasi2021application}.
Nevertheless, running DTs requires significant investments in computational power and storage. Given the superior capabilities of cloud servers, outsourcing these DT inference tasks to professional cloud service providers, i.e., a paradigm known as Machine Learning as a Service (MLaaS), has become a popular trend in both industry and academia \cite{chen2022privdt} \cite{xu2023adaptively} \cite{hu2023content}. This approach enables resource-constrained users and devices to efficiently obtain high-quality prediction services.
% \looseness=-1

However, this practice raises significant privacy concerns. 
Firstly, training a robust DT model involves substantial investments in data collection, computational resources, and specialized expertise. Model providers are often reluctant to expose their models in plaintext on the cloud due to the risk of intellectual property theft.
Secondly, users frequently input sensitive information, such as financial details or medical records, into these models. Transmitting such data unencrypted over the cloud can lead to severe privacy breaches.
Thirdly, users may not want cloud providers to access the actual predictions, particularly in sensitive contexts like financial decisions or medical diagnoses. In such cases, the cloud should execute the prediction without knowing the results or the underlying paths taken for predictions.
To address these concerns, it is essential to embed robust security measures in the framework for outsourcing DT inference to ensure the confidentiality of models, sensitive input data, prediction paths, and final results.
% \looseness=-1

As shown in \autoref{summary}, recent advancements have made notable progress in privacy-preserving DT inference, particularly within the outsourced setting.
Zheng et al. \cite{zheng2020securely} introduce a secure framework for outsourcing DT inference, although it incurred substantial communication overhead on the cloud side. Liu et al. \cite{liu2020towards} develope a comprehensive evaluation scheme that fully protects the privacy of queries, classification results, and decision tree models.
Zhang et al. \cite{zhang2022decision} design an outsourced scheme by combining a modified K-nearest neighbors and one-way functions for DT-based e-Healthcare systems.
However, these methods typically require traversing all decision tree nodes during inference, leading to high computational overhead.
% \looseness=-1

To improve efficiency, recent outsourced research has focused on methods that traverse only the correctly predicted path. 
Ma et al. \cite{ma2021let} develop a protocol that combines the strengths of two existing paradigms to improve efficiency. 
Bai et al. \cite{bai2022scalable} design a sublinear private decision tree evaluation protocol with $O(d)$ communication complexity. 
Yuan et al. \cite{yuan2024efficient} propose a privacy-preserving scheme for outsourced Gradient Boosting Decision Tree (GBDT) inference that requires accessing only the correct path while ensuring path privacy through an obfuscated permutation of multiple trees. Notably, their scheme is effective only for GBDT, i.e., models composed of multiple trees, and it fails to preserve path privacy when applied to a single decision tree.
However, none of these methods support offline users, who must remain online to obtain predictions. A more detailed exploration of these related works is available in Section \ref{related}.
% \looseness=-1 

To address these challenges, we propose \sysname, a secure and efficient privacy-preserving inference scheme for outsourced DT in the cloud. Our scheme traverses only the correct path and supports offline users, thereby significantly reducing both computation and communication costs. To the best of our knowledge, we are the first to apply functional encryption to fully support privacy-preserving DT inference. In addition, we incorporate lightweight additive secret sharing to improve the computational efficiency on the cloud side. Our framework adopts a two-server model, where the computation is delegated to two independent, semi-honest cloud service providers. This non-collusion assumption is highly justified in practice, as these providers often belong to different companies, both of which value their commercial reputation. Discovery of any collusive behavior would severely damage user trust and result in the loss of substantial market share. In fact, this dual-cloud architecture has emerged as a new trend in security design, frequently appearing in application-specific schemes \cite{zheng2021towards} \cite{zhang2022achieving} \cite{hu2023achieving}.
\looseness=-1

\begin{table*}[t]
    \centering
    \caption{Summary of Existing Outsourced PDTE Protocols.}
    \label{summary}
    \vspace{5pt}
    \renewcommand{\arraystretch}{1.5}
    \resizebox{\textwidth}{!}{
    \begin{tabular}{|c|c|c|c|c|c|c|c|c|c|c|}
        \hline
        \hline
        \textbf{Protocol} & 
        \makecell{\textbf{Client} \\ \textbf{Computation}} & 
        \makecell{\textbf{Server} \\ \textbf{Computation}} & 
        \textbf{Round} & 
        \textbf{Leakage} & 
        \textbf{One Path} & 
        \textbf{Off-line} & 
        \makecell{\textbf{Masked} \\ \textbf{Model}} & 
        \makecell{\textbf{Trusted} \\ \textbf{Initializer}} & 
        \textbf{Servers} & 
        \textbf{Techniques} \\
        \hline
        \hline
        \cite{de2017efficient} & $O((m+n)t+d)$ & $O(mt+2^d)$ & $\approx 9$ & $d$ & \Circle & No & No & Yes & Single & ASS \\
        \hline
        \cite{liang2019efficient} & $O(nt)$ & $O(mdt)$ & 1 & $m, d$ & \Circle & Yes & Yes & No & Single & PRF, SKE \\
        \hline
        \cite{liu2020towards} & $O(nt)$ & $O(mt)$ & 1 & $n, \overline{n}, m$ & \Circle & Yes & No & Yes & Two & AHE, ASS \\
        \hline
        \cite{zheng2020securely} & $O((n+2^d)t)$ & $O((n+t)mt)$ & 1 & $m, d$ & \Circle & Yes & Yes & Yes & Two & ASS \\
        \hline
        \cite{ma2021let} & $O((n+t)d)$ & $O(dt+2^d)$ & $2d-1$ & $n, t, d$ & \CIRCLE & No & Yes & No & Single & SS, OT, GC, COT \\
        \hline
        \cite{bai2022scalable}(PRF) & $O((m+n)dt)$ & $O((m+n)dt)$ & $(3r_F+5)d$ & $m, d$ & \CIRCLE & No & Yes & No & Single & SS, OT, PRF \\
        \hline
        \cite{bai2022scalable}(HE) & $O((m+n)d)$ & $O((m+n)d)$ & $8d$ & $m, d$ & \CIRCLE & No & Yes & No & Single & SS, OT, AHE \\
        \hline
        \cite{zhang2022decision} & $O(mt)$ & $O(mt)$ & 1 & $m, d$ & \Circle & Yes & Yes & Yes & Two & KNN \\
        \hline
        \cite{yuan2024efficient} & $O((M+n)t)$ & $O(Mt)$ & $2d$ & $M,d$ & \LEFTcircle & No & Yes & No & Single & PRF, AHE \\
        \hline
        Ours & $O(mt)$ & $O(dt)$ & $1$ & $m, d$ & \CIRCLE & Yes & Yes & Yes & Two & FE, ASS \\
        \hline
    \end{tabular}%
    }
    
    \vspace{5pt}
    \begin{minipage}{\textwidth}
        \small
        \textbf{Round}: server \& user communication rounds, \textbf{One Path}: traverse only the correct path, \textbf{Masked Model}: the model provider delivers the encrypted model to the server. 
        \textbf{ASS}: Additive Secret Sharing,
        \textbf{PRF}: Pseudo-random Function,
        \textbf{SKE}: Symmetric key encryption,
        \textbf{HE}: Homomorphic encryption,
        \textbf{AHE}: Additively homomorphic encryption,
        \textbf{SS}: Secret sharing,
        \textbf{OT}: Oblivious transfer,
        \textbf{GC}: Garbled circuit,
        \textbf{COT}: Conditional oblivious transfer,
        \textbf{KNN}: K-nearest neighbors,
        \textbf{FE}: Functional encryption.
        $m$: the number of tree nodes, $M$: the number of decision trees, $n$: the dimension of a feature vector, $\overline{n}$: the possible value range of every attribute, $d$: the depth of a tree, $t$: the bit size of feature values, $r_F$: the number of rounds required for securely evaluating PRF $F$. \CIRCLE: yes, \Circle: no, \LEFTcircle: partially support.
    \end{minipage}
\end{table*}

The innovation for our work lies in the following three aspects.
Firstly, we ensure the encryption of decision trees while maintaining their functionality for cloud-based inference. We use pseudo-random functions to conceal feature indexes, allowing secure access to comparison data without revealing sensitive information, thus preserving the integrity and privacy of the decision tree.
Secondly, we simplify the inference process by transforming the comparison function into a linear equation and applying functional encryption for secure computation. This enhances both security and computational efficiency. 
Thirdly, we obscure the actual prediction path by evaluating only a subset of internal nodes, which significantly boosts computational efficiency and strengthens privacy.
Our contributions are summarized as follows:

\begin{itemize}
    \item We present a secure and efficient framework for cloud-based decision tree inference that ensures the privacy of models, feature vectors, prediction paths, and results. This framework utilizes a two-cloud architecture and allows for user disconnection without compromising either privacy or performance.
    \looseness=-1

    \item We propose a protocol that utilizes functional encryption to enable secure decision tree inference. Our approach guarantees the confidentiality of decision trees and user data in semi-honest cloud environments, while ensuring efficiency by traversing only the correct prediction path.

    \item We validate the robustness of our security design through empirical evaluations on practical decision tree sizes, achieving online inference in less than 50 milliseconds. Moreover, we compare our approach with existing privacy-preserving schemes, and the experimental results show that our method offers significant advantages in both computational and communication overhead.
\end{itemize}

% Additionally, we provide a comparative analysis with other privacy-preserving schemes, highlighting our framework as the first to support offline users while traversing only the correct prediction path.

\section{Related Works}
\label{related}
Research on privacy-preserving decision tree evaluation (PDTE) has been extensively explored from multiple perspectives, including cryptographic approaches, hardware-assisted solutions, and differential privacy-based methods. Hardware-assisted schemes often incur high costs, while differential privacy-based approaches can compromise model accuracy. 
In contrast, cryptography-based schemes provide strong privacy guarantees without sacrificing correctness, making them highly suitable for secure and practical decision tree evaluation. Consequently, this work focuses on cryptography-based PDTE protocols, which offer strong privacy guarantees and broad applicability.
\looseness=-1

Bost et al. \cite{bost2014machine} model a decision tree as a high-degree polynomial, enabling the server to compute the prediction directly over the client’s FHE-encrypted input through homomorphic operations. To eliminate the reliance on heavy FHE, Wu et al. \cite{wu2015privately} instead require the server to send the decision tree to the client. For security, the server must first convert the tree into a randomized and complete structure before sending it. However, this transformation leads to an exponential increase in the server’s computational cost with respect to the tree depth. 
Tai et al. \cite{tai2017privacy} further improve Wu et al.’s work by cleverly exploiting the structure of decision trees. They showed that their work successfully avoids an exponential number of encryptions in the depth of the decision tree.
Kiss et al. \cite{kiss2019sok} systematically categorize constant-round protocols to identify optimal instantiations, utilizing garbling techniques and homomorphic encryption.
Tueno et al. \cite{tueno2019private} propose a method to represent a decision tree as an array, and implement oblivious array indexing. In their scheme, garbled circuits (GC) \cite{yao1986generate}, oblivious transfer (OT) \cite{naor2001efficient} or oblivious RAM (ORAM) \cite{li2016privacy} is adopted. Specifically, the use of ORAM results in a decision tree method with sublinear complexity of the size of the tree.
Cong et al. \cite{cong2022sortinghat} propose an efficient binary decision tree evaluation technique in the fully homomorphic encryption setting.
Kjamilji et al. \cite{kjamilji2024privacy} design a novel secure and private DT evaluation and its extension over malicious clients protocols.
However, the above approaches rely on a ``server–client'' model in which the server holds the model in plaintext. In this setting, the model provider is equivalent to the server and must perform the classification service locally. As a result, these methods are unsuitable for outsourced inference, a scenario that is both practical and highly relevant in real-world deployments. In many applications, the model provider lacks the computational resources or availability to deliver continuous online services, making it necessary to delegate the inference task to more powerful cloud servers. Addressing this outsourcing problem is precisely the motivation and focus of the scheme proposed in this paper.

Given the limitations of two-party protocols in the outsourced setting, recent research has shifted toward outsourced PDTE schemes, as summarized in \autoref{summary}.
Cock et al. \cite{de2017efficient} propose an outsourced PDTE protocol in a semi-honest model for computing private scoring of decision trees with only modular additions and multiplications.
Liang et al. \cite{liang2019efficient} propose an efficient and secure decision tree classification scheme in the work-flow for cloud-assisted online diagnosis services. Based on decision tree rule induction technique and searchable symmetric encryption, their scheme protects the confidentiality of diagnosis model and user’s data against the semi-trusted cloud server while achieving computational efficiency.
Similar to Bost et al.’s work \cite{bost2014machine}, Liu et al. \cite{liu2020towards} design a privacy-preserving decision tree evaluation scheme that can fully protect the query privacy, the classification result and the decision tree model simultaneously.
Zheng et al. \cite{zheng2020securely} propose a secure framework for outsourcing decision tree inference with huge communication overhead on the cloud side.
Zhang et al. \cite{zhang2022decision} propose an outsourced PDTE scheme for e-Healthcare scenarios. Their method protects both patients’ sensitive physiological data and medical providers’ decision tree structures by combining a modified KNN-based similarity computation with matrix randomization and monotonic one-way functions to obfuscate intermediate results.
However, as shown in \autoref{summary}, all of the above works require traversing all the nodes of a tree, which can cause huge computational and communication overheads.

To address this problem, several studies investigate outsourced PDTE schemes that traverse only the correct path.
Ma et al. \cite{ma2021let} propose a lightweight protocol that outsources the decision tree to two non-colluding servers for classification. Their outsourced extension supports multi-client joint evaluation and is the first to achieve this without using multikey fully-homomorphic encryption.
Bai et al. \cite{bai2022scalable} design a sublinear PDTE protocol with $O(d)$ communication complexity. The core of their construction is a shared oblivious selection functionality, allowing two parties to perform a secret-shared oblivious read operation from an array.
Yuan et al. \cite{yuan2024efficient} propose an efficient and comprehensive PDTE framework for outsourcing GBDT inference utilizing pseudorandom function and additively homomorphic encryption. 
However, this work cannot protect path privacy when dealing with a single decision tree. The path privacy is preserved only in the case of a GBDT model consisting of multiple trees. This is precisely why the ``One Path'' property of \cite{yuan2024efficient} in \autoref{summary}is so distinctive.
However, as shown in \autoref{summary}, all of these works require a user to remain online and interact with a server at all times. This limitation not only increases the user’s dependence on continuous network availability but also reduces the practicality and scalability of the system in real-world deployments.

\begin{table}[t!]
\caption{Notation used}
% \vspace{1em}
\label{tab:notation}
\begin{center}
\setlength{\tabcolsep}{4mm}{
\begin{tabular}{@{}cl@{}}
\toprule
Notations                   & Definition       \\ \midrule
$n$                         & Number of features                \\
$d$                         & Maximum depth of decision tree    \\
$\gamma$                    & Number of internal nodes  \\
$m$                         & Number of leaf nodes \\
$IN_i$                      & The i-th internal node of the decision tree \\                          
                            & under the hierarchical traversal \\
$f_i$                       & Feature name of the i-th internal node\\
$\theta_i$                  & Threshold value of the i-th internal node\\
$sk$                        & Symmetric key \\
$mpk, msk$                  & Public, Secret key of FE \\
$F$                         & Public pseudorandom function (PRF) \\
$S_{MP}$                    & Seed of PRF generated by model provider related to \\ & security parameter \\
$\llbracket x \rrbracket$   & Ciphertext of $x$ encrypted by $mpk$ \\
$e_{f_i, n}$                & column vector in $\mathbb{R}^n$ with all zeros except \\                   
                            & for a $1$ in position $f_i$ \\
$\mathsf{Fs}_i(x)$                    & Feature selection function of the $IN_i$ \\ 
$R_i$                       & Result of boolean test function at layer i \\
$x_i$                       & Data after feature selection of the $IN_i$ \\
$\langle x \rangle^1 , \langle x \rangle^2$  & Additive share of cloud server 1, cloud server 2 \\
\bottomrule
\end{tabular}}
\end{center}
% \vspace{-2em}
\end{table}

\section{Preliminary}
In this section, we introduce preliminary concepts that serve as the basis of our scheme. \autoref{tab:notation} lists the key notations used throughout this paper.

\subsection{Decision Tree}
A decision tree has internal nodes $\{IN_1, IN_2, ..., IN_\gamma\}$, each of which contains feature index and threshold value $(f_i, \theta_i)$.
Each leaf node $LN_i$ in $\{LN_1, LN_2, ..., LN_m\}$ is associated with a label $l_i$ that determines inference results from evaluating a decision tree. %  (value from $\{l_1, l_2, ..., l_L\}$) 
The inference over a decision tree takes as input an $n$-dimensional feature vector, denoted by $x={x_1, x_2,...,x_n}$, and proceeds as follows. 
Based on a feature index $f_i$ in \{1, 2, ..., n\}, it starts with feature selection by assigning a feature $x_{f_i}$ of $x$ to internal node $IN_i$, which is used to compare with the corresponding threshold $th_i$. Then it comes to the evaluation of internal nodes. Starting at the root node $IN_1$, the boolean testing function $B(x_{f_1}, \theta_1)=(x_{f_1} > \theta_1)=R_1$ is evaluated. The result $R_1$ decides whether to next take the left branch ($R_1=0$) or the right branch ($R_1=1$). Such evaluation includes input selection and a boolean testing function that keeps going until a leaf node $LN_i$ is reached. The depth $d$ of a decision tree is defined by the length of the longest path between the root node and a leaf node. Therefore, a decision tree only needs to evaluate $d$ internal nodes. 
In general, decision trees need not be binary or complete. However, all decision trees can be transformed into a complete binary decision tree by increasing the depth of the tree and introducing dummy internal nodes.
Without loss of generality, here we only consider a complete binary decision tree, which is also consistent with previous works \cite{wu2015privately}, \cite{tai2017privacy}, \cite{de2017efficient}, \cite{tueno2019private}.
\looseness=-1

\subsection{Functional Encryption}
Our scheme relies on a functional encryption (FE) scheme \cite{abdalla2015simple} for the inner-product functionality $IP= (Setup, KeyDer, Encrypt, Decrypt)$ based on a public-key encryption (PKE) scheme $\varepsilon  =(Setup, Encrypt, Decrypt)$. The standard definition for FE is given as follows:

\begin{definition}
[PKE-IP FE] Let $\varepsilon=(Setup, Encrypt, De$-$\\crypt)$ be a PKE scheme with linear key homomorphism and linear ciphertext homomorphism under shared randomness. We define our functional encryption scheme for the inner-product functionality $IP=(Setup, KeyDer, Encrypt, Decrypt)$ as follows.

\begin{itemize}
    \item $Setup(1^ \lambda, \ell, B)$ calls $\varepsilon$'s key generation algorithm to generate $\ell$ independent $(sk_1, pk_1)$, ..., $(sk_\ell, pk_\ell)$ pairs, sharing the same public parameters $params$. Here, $B$ denotes the upper bound of each coordinate in the vector. Then, the algorithm sets the functionality's key space $K_\ell$ and message space $X_\ell$ to $M=\{0,...,B-1\}^\ell \subseteq \mathbbm{Z}_q$ and returns $mpk=(params, pk_1,...,pk_\ell)$ and $msk=(sk_1,...,sk_\ell)$.
    \item $KeyDer(msk,\mathbf{y})$ on input master secret key $msk$ and a vector $\mathbf{y}=(y_1,...,y_\ell) \in M$, computes $sk_y$ as an G-linear combination of $(sk_1,...,sk_\ell)$ with coefficients $(y_1,...,y_\ell)$, namely $sk_y=\sum_{i \in [\ell]}y_i \cdot sk_i$.
    \item $Encrypt(mpk, \mathbf{x})$ on input master public key $mpk$ and message $\mathbf{x}=(x_1,...,x_\ell) \in M$, chooses shared randomness $r$ in the randomness space of $\varepsilon$, and computes $ct_0=\varepsilon .C(r)$ and $ct_i=\varepsilon .E(pk_i, x_i;r)$. Then the algorithm returns the ciphertext $Ct=(ct_0, (ct_i)_{i \in [\ell]})$.
    \item $Decrypt(mpk, Ct, sk_y)$ on input master public key $mpk$, ciphertext $Ct=(ct_0, (ct_i)_{i \in [\ell]})$, and secret key $sk_y$ for vector $\mathbf{y}=(y_1,...,y_\ell)$, returns the output of $\varepsilon.Decrypt(sk_y, (ct_0,\prod_{i\in [\ell]}ct_i^{y_i}))=\mathbf{x} \cdot \mathbf{y}$.
\end{itemize}
\end{definition}

\subsection{Additive Secret Sharing}
Additive secret sharing is a kind of widely used secure multiparty computation technology. We refer to a value $x \in \mathbbm{Z}_q$ that is additively shared in the ring $\mathbbm{Z}_q$ as the sum of two values \cite{shamir1979share}. To split an integer $x$ into two additive shares, one can randomly choose an $\langle x \rangle ^A$ from $\mathbbm{Z}_q$ and set $\langle x \rangle ^B \leftarrow x-\langle x \rangle ^A \  mod \ q$.
In the following, we use $\langle x \rangle$ to denote an integer $x$ shared by two parties. Besides, $\langle x \rangle ^A$ and $\langle x \rangle ^B$ are used to denote the additive shares belonging to party A and B, respectively. To reconstruct the value, one of the parties sends its share to the other, and the other one calculates $x= \langle x \rangle ^A + \langle x \rangle ^B$. We denote the reconstruction as $Rec$. Moreover, we only use addition methods (i.e., $Add$) for the additive shares, and do not calculate the multiplication of two additive shares $\langle x \rangle$ and $\langle y \rangle$.
\looseness=-1

\subsection{Pseudorandom Function}
The standard definition for a pseudorandom function is given as follows:

\begin{definition}
Let $F:\{0,1\}^* \times \{0,1\}^* \rightarrow \{0,1\}^*$ be an efficient, length-preserving, keyed function. We say $F$ is a \textit{pseudorandom function} if for all probabilistic polynomial-time distinguishers \textit{D}, there exists a negligible function \textit{negl} such that:
$$
|Pr[D^{F_k(\cdot)}(1^n)=1]-Pr[D^{f_n(\cdot)}(1^n)=1]| \leq negl(n),
$$
where $k \leftarrow \{0,1\}^n$ is chosen uniformly at random and $f_n$ is chosen uniformly at random from the set of functions mapping n-bit strings to n-bit strings.
\end{definition}

Notice that \textit{D} interacts freely with its oracle. Thus, it can ask queries adaptively, choosing the next input based on the previous outputs received. However, since \textit{D} runs in polynomial time, it can only ask a polynomial number of queries. 
Notice also that a pseudorandom function must inherit any efficiently checkable property of a random function. For example, even if $x$ and $x'$ differ in only a single bit, the output $F_k(x)$ and $F_k(x')$ must (with overwhelming probability over the choice of $k$) look completely uncorrelated.

\begin{figure}[t!]
\centering
\centerline{\includegraphics[scale=.35]{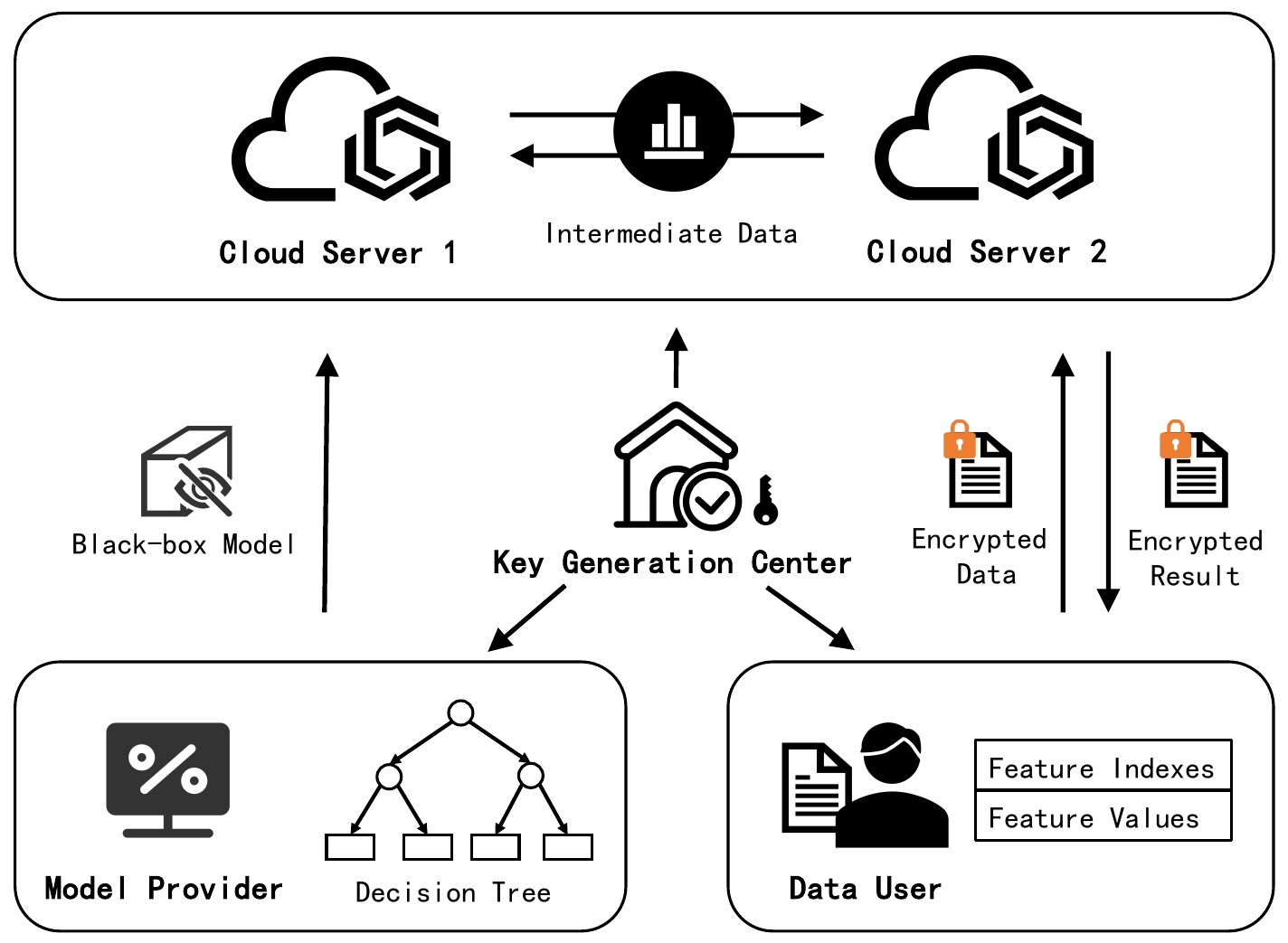}}
\caption{System architecture.}
\label{system}
\end{figure}

\section{Problem Statement}

\subsection{System Architecture}
In this work, we explore a two-cloud architecture model, as depicted in \autoref{system}. Our proposed system involves five types of entities: the Key Generation Center ($KGC$), Model Provider ($MP$), Cloud Server 1 ($CS_1$), Cloud Server 2 ($CS_2$), and Data User ($DU$).

In our system, the $KGC$ generates and distributes public and private keys to all parties. The $MP$ encrypts the complete decision tree and sends it to the cloud servers. The $DU$ splits their sensitive data using additive secret sharing and hides feature indexes with a pseudorandom function before outsourcing the masked data to both $CS_1$ and $CS_2$.
$CS_1$ and $CS_2$ store the encrypted model and manage the $DU$'s data. They collaborate to provide secure decision tree evaluation services, ensuring the $DU$ receives accurate inferences while preserving data privacy and security.

We use a two-cloud architecture to enhance privacy and security. It’s essential that the two cloud servers do not collude. For example, if the $MP$ deploys the service on Amazon AWS, AWS alone cannot ensure full privacy. 
Thus, it partners with other providers, such as Microsoft Azure, to offer secure inference services that also support offline users.
These companies avoid collusion to protect their reputations, as collusion would risk their market standing. The two-cloud approach is widely adopted in privacy-preserving machine learning \cite{zheng2021towards} \cite{zhang2022achieving} \cite{hu2023achieving}, leveraging multiple independent providers to boost security and privacy for various applications.

\subsection{Threat Model}
\label{sec:TM}

Building upon the two-cloud architecture introduced earlier, we focus on the security assumptions and potential behaviors among the involved entities. In our model, the KGC is trusted by all parties and is responsible for generating and distributing public and private keys. 

Our protocol is analyzed under the \textit{semi-honest adversary model} in a stand-alone setting~\cite{mohassel2017secureml, dong2020privacy, adams2021privacy}. That is, adversaries follow the protocol specification but try to infer additional information from received messages. We assume \textit{static corruption}, where at most one of the cloud servers, or the user, may be corrupted. The two cloud servers are assumed to be non-colluding, a realistic assumption supported by their operation under distinct cloud providers~\cite{mohassel2017secureml, cai2018leveraging}.

We aim to protect the following privacy aspects:
\begin{itemize}
    \item \textbf{Input privacy:} The user’s input feature vector $x$ must remain hidden from the cloud servers and the model provider.
    \item \textbf{Output privacy:} The final prediction result $f_T(x)$ must be visible only to the user. Neither the cloud servers nor the model provider should learn the output of the inference.
    \item \textbf{Model privacy:} The structure and parameters of the decision tree $T$ (e.g., feature indexes, thresholds, and leaf labels) must remain hidden from the user and the servers.
    \item \textbf{Prediction path privacy:} No party, including the servers, should learn the full sequence of internal nodes traversed during prediction.
\end{itemize}

We formalize these goals using a \textit{simulation-based security definition}. Informally, a protocol $\Pi$ is secure if whatever a probabilistic polynomial-time (PPT) adversary can observe in the real execution can be simulated in an ideal world where it only learns the allowed outputs. Formally:

\begin{definition}[\textit{Simulation-based Security}]
\textit{Let $\mathsf{Real}_\mathcal{A}(x, T)$ denote the view of a PPT adversary $\mathcal{A}$ in a real execution of the protocol $\Pi$ with user input $x$ and model $T$. Let $\mathsf{Ideal}_\mathcal{S}(f_T(x))$ denote the output of a simulator $\mathcal{S}$ interacting only with the functionality $f_T(x)$. We say $\Pi$ is secure against $\mathcal{A}$ if:}
\[
\mathsf{Real}_\mathcal{A}(x, T) \approx_c \mathsf{Ideal}_\mathcal{S}(f_T(x)),
\]
\textit{where $\approx_c$ denotes computational indistinguishability.}

The above holds separately for:
\begin{itemize}
    \item $\mathcal{A}$ corrupting $CS_1$ or $CS_2$ (input and path privacy),
    \item $\mathcal{A}$ corrupting $DU$ (model privacy).
\end{itemize}
\end{definition}

\begin{figure*}[t!]
\centering
\centerline{\includegraphics[scale=.4]{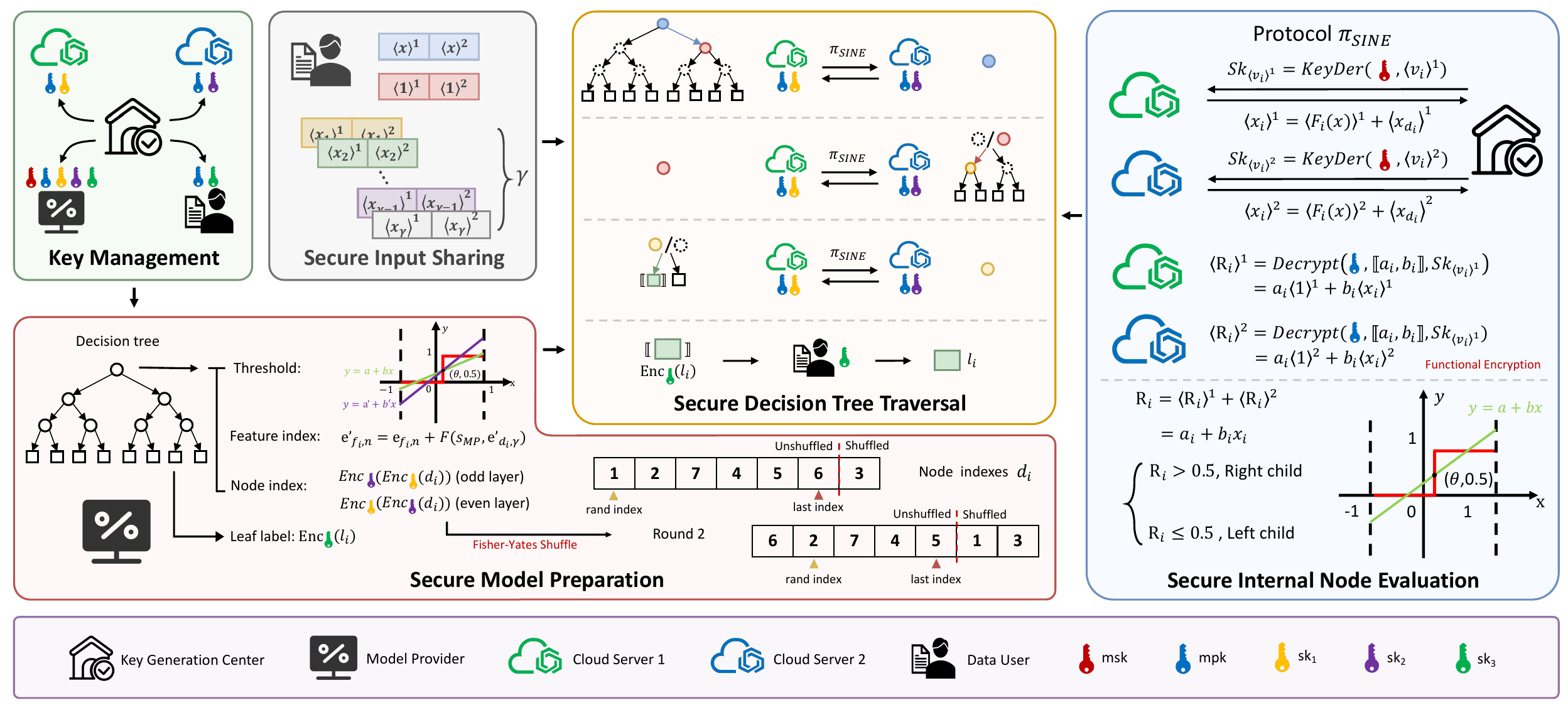}}
\caption{The workflow of our \sysname.}
\label{workflow}
% \vspace{-1.0em}
\end{figure*}

\section{Efficient and Privacy-preserving Decision Tree Inference in the Cloud}
\subsection{Overview}
Our protocol is the first to leverage lightweight functional encryption, specifically tailored to optimize both privacy and overhead in cloud-based decision tree inference.
As shown in \autoref{workflow}, our protocol is comprised of the following phases: (1) key management, (2) secure model preparation, (3) secure input sharing, (4) secure internal node evaluation, and (5) secure decision tree traversal. 
Specifically, the $KGC$ first distributes the cryptographic keys. Before uploading the model, the $MP$ encrypts the information of the internal and leaf nodes, including thresholds, feature indexes, node indexes, and leaf labels. During inference, the $DU$ can go offline after converting the input to secret sharings and sending it to two servers. Note that the two servers securely traverse only the correct decision path without accessing all internal nodes. This is achieved through a secure internal node evaluation protocol, which leverages functional encryption to evaluate internal nodes and determine whether to proceed to the left or right child. The $DU$ can then decrypt the received result to obtain the final prediction.
We give the details in the subsequent sections.

\subsection{Key Management}
Before deploying the model to the cloud for inference, we need to define the key management process to ensure that the subsequent steps can run smoothly.
The $KGC$ uses the key generation algorithm $Setup(1^\lambda, \ell, B)$ to get ($mpk$, $msk$) and sends ($mpk$, $msk$) to the $MP$. 
The $DU$, $CS_1$ and the $CS_2$ receive $mpk$.
Moreover, the $KGC$ generates three different symmetric keys, i.e., $sk_1$, $sk_2$, $sk_3$. 
These keys are then distributed to the relevant parties, with $sk_1$ being sent to the $CS_1$, $sk_2$ being sent to the $CS_2$, and $sk_3$ being sent to the $DU$.
The $KGC$ sends the $sk_1$, $sk_2$, and $sk_3$ to the $MP$.
The purpose of $sk_1$ and $sk_2$ is to avoid revealing the true prediction path when using node indexes. 
In addition, $sk_3$ is used to conceal the prediction results of leaf nodes.
\autoref{workflow} illustrates the key distribution by the $KGC$ to different parties.

\subsection{Secure Model Preparation}
\label{model_preparation}
In outsourced inference scenarios, the $MP$ deploys the decision tree on cloud servers. However, directly uploading the decision tree may expose private information from the training dataset and compromise the $MP$’s ownership of the model. In our protocol, the $MP$ encrypts all node information of the decision tree and can go offline after uploading the secured model. The cloud servers can only perform inference to obtain encrypted prediction results, without learning any internal information about the decision tree or the actual prediction path. Next, we provide a detailed description of the operations performed by the $MP$ during the secure model preparation phase.
\looseness=-1

As shown in \autoref{workflow}, we assume that the $MP$ owns a complete binary tree $\mathbf{T}$ of depth $d$. This tree contains $2^{d+1}-1$ nodes, where $m = 2^{d}$ are leaf nodes and $\gamma = 2^{d}-1$ are internal nodes. 
In our algorithm $\pi_{SMP}$ in \autoref{Algorithm1}, the $MP$ first encrypts a label with $sk_3$ for each leaf node $\mathbf{LN}_i$ (where $i = 1, 2, \cdots, m$), 
resulting in $Enc_{sk_3}(l_i)$. Consequently, the two-cloud can only get encrypted leaf values, while the $DU$ can decrypt final predicted values by $sk_3$.

Next, we consider the information of internal nodes as steps 2 to 5 in \autoref{Algorithm1}.
Each internal node $IN_i$ contains feature index $f_i$ and threshold value $\theta_i$.
The feature index $f_i$ is an index in \{1, 2, ..., $n$\}, and each internal node $IN_i$ identifies a data $x_i$ used for a boolean test function with a function $\mathsf{Fs}_i(x) = e^T_{f_{i, n}} \cdot x$, where $\cdot$ is the standard row-by-column multiplication. Here, the vector $e_{f_i, n}$ is a column vector in $\mathbb{R}^n$ with all zeros except for a 1 in position $f_i$, and $e^T_{f_i, n}$ is its transpose.
In order to hide the information of $f_i$ and to keep the function $\mathsf{Fs}_i$ available, we first designed the following reindexing system for a complete binary tree. 
Assume an array $A=[1,2, \cdots, \gamma]$, we use the Fisher-Yates shuffle algorithm \cite{fisher1953statistical} to disrupt the array $A$ as step 2 in $\pi_{SMP}$.
Similar to \autoref{workflow}, we iterate $A$ from the end to the beginning (or vice versa). For each location $i$, we swap the value at $i$ with the value at a random target location $j$. Finally, we obtain a disordered array $D=[d_1, d_2,..., d_\gamma]$.
The Fisher-Yates shuffle algorithm requires very few steps, and each iteration requires only a random integer and a swap operation.

\begin{figure}[t]
\centering
\small
\begin{tabular}{|p{8.0cm}|}
\hline \\
 \textbf{Preamble:} Consider a complete binary tree $\mathbf{T}$ consisting of $m$ leaf nodes and $\gamma$ internal nodes. Let the leaf nodes be $\mathbf{LN}_1, \mathbf{LN}_2, \cdots, \mathbf{LN}_m$, and the internal nodes be $\mathbf{IN}_1, \mathbf{IN}_2, \cdots, \mathbf{IN}_\gamma$. \\
 \textbf{Input:} $MP$ holds a complete binary tree $\mathbf{T}$, $(mpk, msk)$, symmetric keys $sk_1$, $sk_2$, and $sk_3$. \\
 \textbf{Output:} $CS_1$ obtains an encrypted decision tree $\mathbf{E_T}$, and $CS_2$ obtains an encrypted root node $\mathbf{E_{IN_1}}$.\\
 \textbf{Procedure}:
 \begin{packeditemize}
    \item[1.] For all leaf nodes $\mathbf{LN}_1, \mathbf{LN}_2, \cdots, \mathbf{LN}_m$, $MP$ encrypts the label $l_i$ of each node $\mathbf{LN}_i$ with $sk_3$: $Enc_{sk_3}(l_i)$.
    \item[2.] $MP$ disrupt the array $A = [1, 2, \cdots, \gamma]$ with the Fisher-Yates shuffle algorithm \cite{fisher1953statistical} to get the array $D = [d_1, d_2, \cdots, d_\gamma]$. Then $MP$ assigns the disrupted array $D$ to internal nodes in hierarchical traversal as new node indexes.
    \item[3.] $MP$ encrypts all node indexes. For odd layers, $MP$ encrypts each node index $d_i$ with $sk_1$ followed by $sk_2$, i.e., $Enc_{sk_2}(Enc_{sk_1}(d_i))$. For even layers, $MP$ encrypts each node index $b_i$ with $sk_2$ followed by $sk_1$, i.e., $Enc_{sk_1}(Enc_{sk_2}(d_i))$. Note the special case where the root node index received by $CS_1$ is encrypted using $sk_1$, i.e., $Enc_{sk_1}(d_1)$, while the root node received by $CS_2$ is encrypted using $sk_2$, i.e., $Enc_{sk_2}(d_1)$.
    \item[4.] $MP$ randomly samples a seed $s_{MP} \leftarrow \{0,1\}^\gamma$ for a public pseudorandom function $F: \{0,1\}^\gamma \times \mathbb{Z}^2 \rightarrow \{0,1\}^n$ and sends the encrypted seed $Enc_{sk_3}(s_{MP})$ to $KGC$.
    \item[5.] For each $\mathbf{IN_i}$,
    \begin{packeditemize}
    \item $MP$ masks \textbf{feature index $f_i$} with $F$ and node index $d_i$: $e_{f_i,n}'=e_{f_i, n} + F(s_{MP}, e_{d_i, \gamma})$. 
    Here, the vector $e_{f_i, n}$ is the column vector in $\{0,1\}^n$ with all zeros except for a 1 in position $f_i$.
    \item To mask \textbf{threshold $\theta_i$}, $MP$ converts a boolean test function into a linear function \( y = a + b x \). Specifically, $MP$ randomly selects an integer \( b \in (1,100) \) and finds \( a \) based on the point \( (\theta, 0.5) \).
    % To mask \textbf{threshold $\theta_i$}, $MP$ converts a boolean test function into a linear function $y=a+bx$. Specifically, $MP$ randomly selects an integer $b \in (1,100)$ and finds $a$ based on the point $(\theta, 0.5)$.\footnote{xxx} 
    \item $MP$ encrypts coefficients with $mpk$: $\llbracket a_i, b_i \rrbracket = Encrypt(mpk, [a_i, b_i])$.
    \end{packeditemize}
    \item[6.] $MP$ sends the encrypted decision tree $\mathbf{E_T} = \{\mathbf{E_{IN_1}}, \cdots, \mathbf{E_{IN_\gamma}}; \mathbf{E_{LN_1}}, \cdots, \mathbf{E_{LN_m}}\}$ to $CS_1$. The encrypted root node $\mathbf{E_{IN_1'}} = \{e_{f_1,n}', \llbracket a_1, b_1 \rrbracket, Enc_{sk_2}(d_1)\}$ is sent to $CS_2$.
    \end{packeditemize} \\
\hline
\end{tabular}
% \vspace{2pt}
\caption{Algorithm $\pi_{SMP}$ for secure model preparation.}
\label{Algorithm1}
% \vspace{-15pt}
\end{figure}

Next, we assign indexes to internal nodes in hierarchical traversal according to the array $D$. The root node is the first internal node, $IN_1$, setting its node index to $d_1$. Then we label the remaining internal nodes level by level and in order from left to right.
The hierarchical order of these node indexes is the same as $D$, i.e., node indexes of internal nodes are disrupted.
Moreover, the $MP$ encrypts node indexes of even and odd layers using both $sk_1$ and $sk_2$.
Note that the index of a root node received by $CS_1$ is encrypted by $sk_1$. 
$CS_2$ receives an index encrypted by $sk_2$.
As shown in step 3 in \autoref{Algorithm1}, the indexes at even layers are encrypted first with $sk_2$ and then with $sk_1$, and the indexes at odd layers are encrypted first with $sk_1$ and then with $sk_2$.
Specifically, the node indexes of the second layer are encrypted with $sk_2$ and $sk_1$ i.e., $Enc_{sk_1}(Enc_{sk_2}(d_2))$ and $Enc_{sk_1}(Enc_{sk_2}(d_3))$, the node indexes of the third layer is encrypted using $sk_1$ and $sk_2$, e.g., $Enc_{sk_2}(Enc_{sk_1}(d_4))$, $Enc_{sk_2}(Enc_{sk_1}(d_5))$, etc.
In general, $CS_1$ and $CS_2$ can only get the indexes of internal nodes on the prediction path.

In step 4, we assume that $F: \{0,1\}^\gamma \times \mathbb{Z}^2 \rightarrow \{0,1\}^n$ is a public pseudorandom function (PRF). 
The $MP$ samples a seed $s_{MP} \leftarrow \{0,1\}^\gamma$ and sends an encrypted seed $Enc_{sk_3}(s_{MP})$ to the $KGC$.
As shown in step 5 in \autoref{Algorithm1}, for $i = 1,2, \cdots, \gamma$, the $MP$ computes $e_{f_i,n}'=e_{f_i, n} + F(s_{MP}, e_{d_i, \gamma})$. Therefore, the $MP$ hides each feature index $f_i$ by adding the output of PRF on the vector $e_{f_i, n}$. 
After these operations, the feature index of each internal node $f_i$ becomes a vector $e_{f_i,n}'$. 
As a result, the boolean test function is $\mathsf{Fs}_i(x)={e'}_{f_i,n}^T \cdot x$.
With no $s_{MP}$, the adversary cannot derive information about $f_i$ based on $e_{f_i,n}'$.
\looseness=-1

After hiding feature indexes of internal nodes, we now process the threshold $\theta_i$ for each internal node. 
Unlike other schemes that encrypt thresholds, we choose to convert a boolean test function into a linear function. 
The essence of the threshold $\theta$ is to define a boolean test function $B$ that results in 1 if $x > \theta$ and 0 if $x \leq \theta$. 
This is a non-linear comparison function. 
Observing the function $B$ on the x-y coordinate axis in \autoref{workflow}, note that the point $P = (\theta, 0.5)$, we set the linear function $y=a+bx$, $b>0$, and the function $y$ passes through the point $P$. 
For simplicity, the $MP$ randomly selects an integer $b \in (1,100)$ and finds $a$ based on the point $(\theta, 0.5)$.\footnote{To ensure all operations are performed in integer space, $MP$ scales the values in the formula by a factor of \( 10^k \) (for some integer \( k \)), where \( k \) is chosen such that the decimal parts are eliminated. Specifically, the formula \( b = 0.5 + a * \theta \) is transformed into \( \tilde{b} = \lfloor (0.5 + a * \theta) \cdot 10^k \rfloor \), where \( \lfloor \cdot \rfloor \) denotes the floor function. This scaling and truncation process ensures that all values involved in the encryption process are integers, making subsequent operations compatible with integer-based cryptographic schemes.}
If the two-cloud can compute a plaintext result $R$ of the linear function, it only takes comparing $R$ with $0.5$ to get a result $R=0$ or $1$ of the corresponding boolean test function, i.e., to know which child node is selected next.
Therefore, for each internal node $IN_i$ with threshold $\theta_i$, the $MP$ converts a boolean test function into a linear function to obtain the coefficients $a_i$ and $b_i$. Then, the $MP$ uses $mpk$ to encrypt coefficients, so the threshold for each internal node becomes $\llbracket a_i, b_i \rrbracket = Encrypt(mpk, [a_i, b_i])$.

Finally, the information of both internal and leaf nodes in a decision tree $\mathbf{T}$ is encrypted or disturbed.
As shown in \autoref{Algorithm1}, the $MP$ sends an encrypted decision tree $\mathbf{E_T}$ to $CS_1$ and an encrypted root node $\mathbf{E_{IN_1}}$ to $CS_2$. Note that the $CS_2$ can only receive information about the root node, i.e., $f_1'$, $d_1$, and $\llbracket a_1, b_1 \rrbracket$. 
\looseness=-1

\subsection{Secure Input Sharing}
\label{input_shares}
In our protocol, the user protects the input through secret sharing and generates additional information to assist with the inference of the decision tree. Specifically, given the feature vector $x$, the $DU$ chooses a random vector $r$ with the same size as $x$. Each element in $r$ is randomly chosen from the rang $\mathbb{Z}_{2^l}$, where $l$ is a sufficiently large parameter (typically 32 or 64 \cite{ziegeldorf2017privacy}). The ciphertexts of $x$ are produced as $\langle x \rangle^1 = \{(x_i-r_i) \ mod \ 2^l\}_{i=1}^n$ and $\langle x \rangle^2 = \{r_i \ mod \ 2^l\}_{i=1}^n$.
The $\langle x \rangle^i$ is sent to the two-cloud $CS_i$ ($i \in \{1,2\}$). 
Note that we have $x = (\langle x \rangle^1 + \langle x \rangle^2) \ mod \ 2^l$ through element-wise addition. 
Similarly, the $DU$ also encrypts $1$ as $\langle 1 \rangle^1$ and $\langle 1 \rangle^2$, and sends them to $KGC$.

Furthermore, the $DU$ receives $Enc_{sk_3}(s_{MP})$ with $sk_3$ from $KGC$ and gets the real value of $s_{MP}$. 
For $j=1, 2, ..., \gamma$, the $DU$ computes $\langle x_j \rangle^1 = -F(s_{MP}, e_{j, \gamma})^T \cdot \langle x \rangle^1$ and $\langle x_j \rangle^2 = -F(s_{MP}, e_{j, \gamma})^T \cdot \langle x \rangle^2$. Afterwards, $DU$ sends them to $CS_1$ and $CS_2$ respectively.  
These secret shares, such as $\langle x_j \rangle^1$ and $\langle x_j \rangle^2$, are used to perform the boolean test function of internal nodes, which we will describe next in detail. After sending the secret shares to the cloud servers, the $DU$ can go offline and does not need to remain online during the inference phase.

\subsection{Secure Internal Node Evaluation}
\label{sine}

\begin{figure}[t]
\centering
\small
\begin{tabular}{|p{8.0cm}|}
\hline \\
 \textbf{Input:} $CS_1$ holds an encrypted node $\mathbf{E_{IN_i}}$, an additive secret share $\langle x \rangle^1$, and additive secret shares $\langle x_j \rangle^1$, j = $1, 2, \cdots, \gamma$. 
 $CS_2$ holds an encrypted node $\mathbf{E_{IN_i}}$, an additive secret share $\langle x \rangle^2$, and additive secret shares $\langle x_j \rangle^2$, j = $1, 2, \cdots, \gamma$. 
 $KGC$ holds $\langle 1 \rangle^1$ and $\langle 1 \rangle^2$.\\
 \textbf{Output:} $CS_1$ obtains a comparison result $\mathbf{R}$.\\
 \textbf{Procedure}:
 \begin{packeditemize}
    \item[1.] The encrypted node $\mathbf{E_{IN_i}}$ includes a masked feature vector $e_{f_i,n}'$, an encrypted threshold $\llbracket a_i, b_i \rrbracket$, and a node index $d_i$.
    \item[2.] $CS_1$ computes the feature selection function: $\langle x_i \rangle^1 = \langle \mathsf{Fs}_i(x) \rangle^1 + \langle x_{d_i} \rangle^1 = {e'}_{f_i,n}^T \cdot \langle x \rangle^1 + \langle x_{d_i} \rangle^1 = e_{f_i, n}^T \cdot \langle x \rangle^1$. Then $CS_1$ sends $\langle x_i \rangle^1$ to $KGC$.
    \item[3.] $CS_2$ computes the feature selection function: $\langle x_i \rangle^2 = \langle \mathsf{Fs}_i(x) \rangle^2 + \langle x_{d_i} \rangle^2 = {e'}_{f_i,n}^T \cdot \langle x \rangle^2 + \langle x_{d_i} \rangle^2 = e_{f_i, n}^T \cdot \langle x \rangle^2$. Then $CS_2$ sends $\langle x_i \rangle^2$ to $KGC$.
    \item[4.] For data $\langle v_i \rangle^1 = [\langle 1 \rangle^1, \langle x_i \rangle^1]$, $KGC$ generates the corresponding $Sk_{\langle v_i \rangle^1}$ with $KeyDer(msk, \langle v_i \rangle^1)$ and sends the $Sk_{\langle v_i \rangle^1}$ to $CS_1$.
    \item[5.] For data $\langle v_i \rangle^2 = [\langle 1 \rangle^2, \langle x_i \rangle^2]$, $KGC$ generates the corresponding $Sk_{\langle v_i \rangle^2}$ with $KeyDer(msk, \langle v_i \rangle^2)$ and sends the $Sk_{\langle v_i \rangle^2}$ to $CS_2$.
    \item[6.] $CS_1$ invokes $Decrypt(mpk, \llbracket a_i, b_i \rrbracket, Sk_{\langle v_i \rangle^1})$ to get the inner product of vector $[a_i, b_i]$ and vector $[\langle 1 \rangle^1, \langle x_i \rangle^1]$. The result is noted as $\langle R_i \rangle^1= a_i\langle 1 \rangle^1 + b_i\langle x_i \rangle^1$.
    \item[7.] $CS_2$ invokes $Decrypt(mpk, \llbracket a_i, b_i \rrbracket, Sk_{\langle v_i \rangle^2})$ to get the inner product of vector $[a_i, b_i]$ and vector $[\langle 1 \rangle^2, \langle x_i \rangle^2]$. The result is noted as $\langle R_i \rangle^2= a_i\langle 1 \rangle^2 + b_i\langle x_i \rangle^2$ and is sent to $CS_1$ (or $\langle R_i \rangle^1$ is sent to $CS_2$).
    \item[8.] $CS_1$ (or $CS_2$) calculates the evaluation result $R_i = \langle R_i \rangle^1+ \langle R_i \rangle^2 = a_i + b_ix_i$. Then $CS_1$ (or $CS_2$) compares $R_i$ with 0.5. If $R_i>0.5$, then the right child node is selected. Otherwise, the left child node is selected.
    \end{packeditemize} \\
\hline
\end{tabular}
% \vspace{2pt}
\caption{Our protocol $\pi_{SINE}$ for secure internal node evaluation.}
\label{Protocol1}
% \vspace{-15pt}
\end{figure}

Given the shares of feature vectors from the $DU$ and the $MP$'s encrypted decision tree, $CS_1$ and $CS_2$ can perform decision tree inference with the assistance of the $KGC$, without requiring the $MP$ or $DU$ to remain online. 
Before presenting the complete inference protocol, we first introduce the evaluation of an internal node to determine the child node.
The evaluation of an internal node involves two steps: feature selection and threshold comparison. Specifically, in the plaintext setting, the internal node typically selects the relevant input feature based on its feature index and compares it with a threshold to decide whether to proceed to the left or right child. With our protocol $\pi_{SINE}$, neither of the two servers can infer the input, the feature index or the threshold of the node.

We now first introduce in detail how to determine a feature value for an internal node $IN_i$. 
For simplicity, we will leave out secret sharing for now. An internal node $IN_i$ in the cloud has a feature index $e'_{f_i,n}$ and a node index $d_i$. 
As a result, the feature selection function becomes $\mathsf{Fs}_i(x)={e'}_{f_i,n}^T \cdot x=(e_{f_{i,n}}^T + F(s_{MP}, e_{d_i, \gamma}^T)) \cdot x$. As we mentioned in the secure input sharing phase, the cloud server gets $x_{d_i}=-F(s_{MP}, e_{d_i, \gamma}^T) \cdot x$. 
The output of the PRF added by the $MP$ is offset by $DU$'s input. The cloud then computes $x_i = \mathsf{Fs}_i(x) + x_{\alpha_i} = (e_{f_i, n}^T \cdot x)$ to get the feature value corresponding to the true feature index $f_i$ of $IN_i$. 
As shown in \autoref{Protocol1}, for an internal node $\mathbf{E_{IN_i}}$, $CS_1$ computes $\langle x_{i} \rangle^1=\mathsf{Fs}_i(\langle x \rangle^1)+ \langle x_{d_i} \rangle^1=(e^T_{f_{i,n}} + F(s_{MP}, e_{d_i, \gamma})^T) \cdot \langle x \rangle^1 - F(s_{MP}, e_{d_i, \gamma})^T \cdot \langle x \rangle^1 = e^T_{f_{i,n}} \cdot \langle x \rangle^1$ to get the feature sharing value of the corresponding feature index of $IN_i$. Similarly, the $CS_2$ calculates the result $\langle x_{i} \rangle^2 = e^T_{f_{i,n}} \cdot \langle x \rangle^2$. 
Since we use secret sharing, neither $CS_1$ nor $CS_2$ can get the true feature values.
The two-cloud then sends $\langle x_{i} \rangle^1$ and $\langle x_{i} \rangle^2$ to the $KGC$.

Having introduced the secure feature selection operations, we now describe how secure threshold comparison is performed. 
Recall that our construction of model preparation leads to an internal node $IN_i$ having $\llbracket a_i, b_i \rrbracket$, which is shared by $CS_1$ and $CS_2$.
As shown in \autoref{Protocol1}, for $\langle v_i \rangle^1 =[\langle 1 \rangle^1, \langle x_{i} \rangle^1]$, the $KGC$ uses $KeyDer(msk, \langle v_i \rangle^1)$ to generate the corresponding $Sk_{\langle v_i \rangle^1}$. Similarly, for $\langle v_i \rangle^2=[\langle 1 \rangle^2, \langle x_{i} \rangle^2]$, the corresponding $Sk_{\langle v_i \rangle^2}$ is generated. 
Now the $Sk_{\langle v_i \rangle^1}$ and $Sk_{\langle v_i \rangle^2}$ are sent to $CS_1$ and $CS_2$ respectively.
With $mpk$, $\llbracket a_i, b_i \rrbracket$ and $Sk_{\langle v_i \rangle^1}$, the $CS_1$ invokes $Decrypt$ algorithm to get the inner product of vector $[a_i, b_i]$ and vector $[\langle 1 \rangle^1, \langle x_{i} \rangle^1]$, i.e., $\langle R_i \rangle^1=a_i\langle 1 \rangle^1 + b_i \langle x_{i} \rangle^1$. Similarly, the $CS_2$ gets the $\langle R_i \rangle^2=a_i\langle 1 \rangle^2 + b_i \langle x_{i} \rangle^2$ and sends $\langle R_i \rangle^2$ to the $CS_1$. The $CS_1$ calculates the evaluation result $R_i=\langle R_i \rangle^1+\langle R_i \rangle^2 = a_i\langle 1 \rangle^1 + a_i\langle 1 \rangle^2 + b_i \langle x_{i} \rangle^1 + b_i \langle x_{i} \rangle^2 = a_i+b_ix_i$. 
The $CS_1$ only needs to compare $R_i$ with $0.5$; if $R_i>0.5$, the right child node is selected, and vice versa for the left branch.
Note that in some cases $CS_1$ needs to send $\langle R_i \rangle^1$ to $CS_2$, who determines $R_i$, as we will explain in detail in Section \ref{traversal}.
A note on the correctness of $R_i$ is in Section \ref{correctness}.
We emphasize that both $CS_1$ and $CS_2$ know the indexes of internal nodes on the prediction path.
The operations in $\pi_{SINE}$ will only target the internal nodes on the prediction path. 
Next, we will describe how to traverse the decision tree to obtain the prediction result while preserving the privacy of the actual prediction path.

\begin{figure}[t]
\centering
\centerline{\includegraphics[scale=.26]{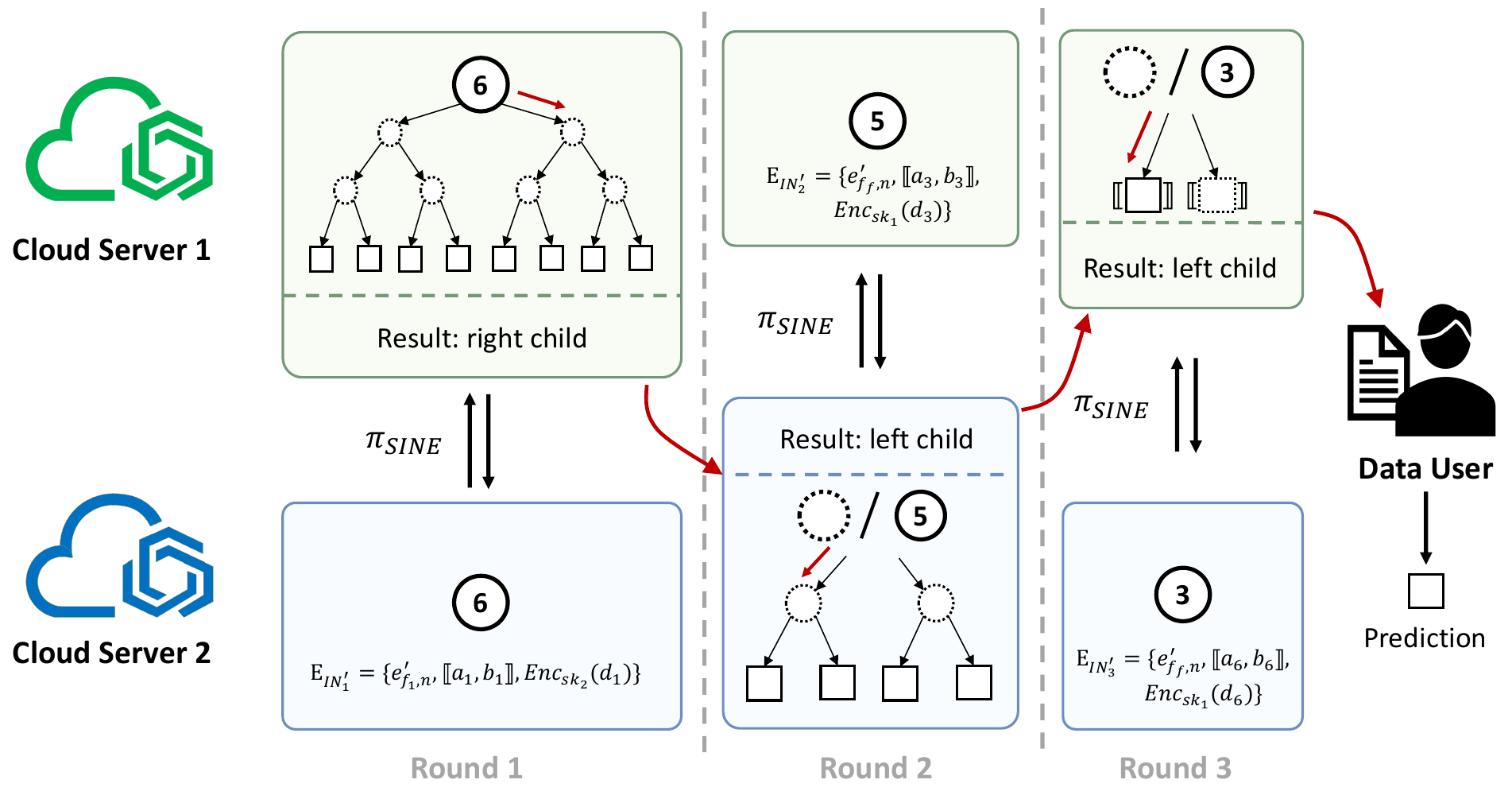}}
\caption{Secure decision tree inference example.}
\label{example}
% \vspace{-2em}
\end{figure}
% \vspace{-1em}

\subsection{Secure Decision Tree Traversal}
\label{traversal}
\begin{figure}[!htp]
\centering
\small
\begin{tabular}{|p{8.0cm}|}
\hline \\
 \textbf{Input:} $CS_1$ holds an encrypted decision tree $\mathbf{E_T} = \{\mathbf{E_{IN_1}}, \cdots, \mathbf{E_{IN_\gamma}}; \mathbf{E_{LN_1}}, \cdots, \mathbf{E_{LN_m}}\}$, an additive secret share $\langle x \rangle^1$, and additive secret shares $\langle x_j \rangle^1$, j = $1, 2, \cdots, \gamma$. 
 $CS_2$ holds an encrypted root node $\mathbf{E_{IN_1}} = \{e_{f_1,n}', \llbracket a_1, b_1 \rrbracket, Enc_{sk_2}(d_1)\}$, an additive secret share $\langle x \rangle^2$, and additive secret shares $\langle x_j \rangle^2$, j = $1, 2, \cdots, \gamma$. 
 $KGC$ holds $msk$, $\langle 1 \rangle^1$, and $\langle 1 \rangle^2$.\\
 \textbf{Output:}  $DU$ obtains an inference result $l_i$.\\ 
 \textbf{Procedure}:
 \begin{packeditemize}
    \item[1.] For the root node (in the first layer), both $CS_1$ and $CS_2$ can obtain the $d_1$. Then, $CS_1$ and $CS_2$ together execute the protocol $\pi_{SINE}$. $CS_1$ gets the evaluation result $R_1$.
    \item[2.] In the second layer:
    \begin{packeditemize}
    \item $CS_1$ selects the subtree $\mathbf{E_{T_2}}$ based on $R_1$ and decrypts the index of the root node in $\mathbf{E_{T_2}}$. Then, $CS_1$ sends the $\mathbf{E_{T_2}}$ to $CS_2$. For example, $CS_1$ selects the left child node based on $R_1$, which is indexed as $Enc_{sk_2}(d_2)$ after decryption.
    \item $CS_2$ decrypts the index of the root node in $\mathbf{E_{T_2}}$ and sends the index to $CS_1$.
    \item $CS_1$ and $CS_2$ execute the protocol $\pi_{SINE}$ together. And $CS_2$ gets the evaluation result $R_2$.
    \end{packeditemize}
    \item[3.] In the third layer:
    \begin{packeditemize}
    \item $CS_2$ selects the subtree $\mathbf{E_{T_3}}$ based on $R_2$, and decrypt the index of the root node in $\mathbf{E_{T_3}}$. Then, $CS_2$ sends the $\mathbf{E_{T_3}}$ to $CS_1$. For example, $CS_2$ selects the right child node based on $R_2$, which is indexed as $Enc_{sk_1}(d_5)$ after decryption.
    \item $CS_1$ decrypts the index of the root node in $\mathbf{E_{T_3}}$ and sends the index to $CS_2$.
    \item $CS_1$ and $CS_2$ execute the protocol $\pi_{SINE}$ together. And $CS_1$ gets the evaluation result $R_3$.
    \end{packeditemize}
    \item[4.] For layer $\in [4, d]$:
    \begin{packeditemize}
    \item For an even layer, $CS_1$ and $CS_2$ perform operations similar to the second layer.
    \item For an odd layer, $CS_1$ and $CS_2$ perform operations similar to the third layer.
    \end{packeditemize}
    \item[5.] After $CS_1$ (or $CS_2$) obtains the evaluation result $R_d$ for layer $d$, $CS_1$ (or $CS_2$) selects the leaf node $\mathbf{E_{LN_i}}$ based on $R_d$ and sends its label $Enc_{sk_3}(l_i)$ to $DU$.
    \item[6.] $DU$ decrypts $Enc_{sk_3}(l_i)$ with $sk_3$ to get the inference result $l_i$.
    \end{packeditemize} \\
\hline
\end{tabular}
% \vspace{2pt}
\caption{Our protocol $\pi_{SDTT}$ for secure decision tree traversal.}
\label{Protocol2}
% \vspace{-15pt}
\end{figure}

With the evaluation results at internal nodes, we now describe how to securely generate the inference result. 
The protocol $\pi_{SINE}$ ensures that neither server can infer the input, the feature indexes, and the threshold values of internal nodes.
Since we utilize the properties of FE to obtain the evaluation results of internal nodes on the prediction path, we have to prevent the prediction path leakage while generating the inference results, and neither $CS_1$ nor $CS_2$ can derive the correct prediction path.
\looseness=-1

We design the secure decision tree traversal protocol $\pi_{SDTT}$ to achieve the above objective.
The process of $\pi_{SDTT}$ is as shown in \autoref{Protocol2}. 
At step 1, $CS_1$ and $CS_2$ perform protocol $\pi_{SINE}$ on the root node of a decision tree, and $CS_1$ gets the true evaluation result $R_1$.
If the result $R_1>0.5$, $CS_1$ selects the right subtree and conversely, the left subtree. The $CS_1$ decrypts the index of the first node of the subtree $\mathbf{E_{T_2}}$ and subsequently sends this subtree $\mathbf{E_{T_2}}$ to $CS_2$. The $CS_2$ decrypts the index of the root node and sends this index to the $CS_1$. Then the two-cloud performs $\pi_{SINE}$ on this node and the $CS_1$ sends the result $R_1$ to the $CS_2$ to finally get the result $R_2$. At this point, $CS_2$ knows which subtree to select next.
\looseness=-1

Similarly, the $CS_2$ decrypts the index of the root node of the subtree $\mathbf{E_{T_3}}$ and sends the subtree to the $CS_1$, which decrypts the index and sends it to the $CS_2$. Then the $CS_1$ leads the node evaluation and cooperates with $CS_2$ to get the evaluation result. 
As shown in step 4 in \autoref{Protocol2}, $CS_1$ and $CS_2$ perform operations similar to step 2 for nodes in even layers. Nodes in odd layers mimic the calculation of step 3.
$CS_1$ and $CS_2$ continue to iterate until the leaf node is reached and send the leaf value to the $DU$. 
Note that the leaf value is encrypted and the $DU$ can get the true prediction result with $sk_3$.

We give a specific example as follows.
As shown in \autoref{example}, the $CS_1$ chooses the right subtree and decrypts the index of the root node of the left subtree, i.e., $Dec_{sk_1}(Enc_{sk_1}(Enc_{sk_2}(d_3)))=Enc_{sk_2}(5)$. Now the $CS_1$ sends the right subtree to $CS_2$. The $CS_2$ decrypts the index of the root node and sends it ($5$) to $CS_1$. Then the two-cloud performs $\pi_{SINE}$ on this node, and the $CS_1$ sends the result $\langle R_2 \rangle^1$ to $CS_2$. The determination result $R_2$ is obtained by the $CS_2$, which selects the right subtree based on $R_2$ and decrypts its root index. The $CS_1$ receives the left subtree and sends plain index $3$ to $CS_2$. After the secure evaluation of the node, the $CS_1$ selects the corresponding leaf node and sends it to the $DU$. The $DU$ uses $sk_3$ to get the true prediction result.

Now we explain why the scheme preserves the privacy of the prediction paths. Firstly, the information of internal nodes, including encrypted feature indexes and thresholds, is indistinguishable, and the only thing two-cloud can know is the index of internal nodes on the prediction path.
Secondly, due to the properties of functional encryption, a server must know the direction of the current node, i.e., we cannot avoid partial leakage of the path. So the core of our scheme is that the server that knows the path of this round does not know the direction of the next round.
As shown in \autoref{example}, the $CS_1$ knows that the right child node is chosen in the first round, but the $CS_1$ does not know where to go in the second round. Even if it knows where the third round is going, it cannot infer the true predicted path. Similarly, the $CS_2$ does not know the direction of the first round, so the direction of the second round does not help infer the true path. By carefully designing the computation protocol between the two servers, we successfully hide the true prediction path.
Overall, our protocol protects the model information of $MP$, the feature vector of $DU$, and hides the prediction path while ensuring the prediction goes smoothly and protecting the prediction results.

\section{Theoretical Analysis}
\subsection{Correctness}
\label{correctness}
% In this section, we prove the correctness of our secure internal node evaluation, i.e., the final result $R$ shows the comparison result of the boolean testing function for node $IN_i$.

For internal node $IN_i$, the two-cloud has encrypted data corresponding to the transformed threshold, i.e., $\llbracket a_i, b_i \rrbracket$.
The $CS_1$ has query shares $\langle 1 \rangle^1$, $\langle x \rangle^1$, and $\langle x_j \rangle^1$, j=$\{1,...,\gamma\}$.
Similarly, the $CS_2$ has secret shares $\langle 1 \rangle^2$, $\langle x \rangle^2$, and $\langle x_j \rangle^2$, j=$\{1,...,\gamma\}$.
Before we perform the proof of correctness, it is important to emphasize the workflow of FE. 
For example, for vector $a = [1, 2]$ and vector $b = [3, 4]$, if vector $a$ is encrypted with FE and vector $b$ generates the corresponding $sk_b$, the result of the inner product of $a$ and $b$ can be obtained directly using FE decryption, i.e., $1*3+2*4 = 11$.

As we mentioned in Section \ref{sine}, for the internal node $\mathbf{IN_i}$ in the protocol $\pi_{SINE}$, $CS_1$ computes $\langle x_{i} \rangle^1=\mathsf{Fs}_i(\langle x \rangle^1)+ \langle x_{\alpha_i} \rangle^1=(e^T_{f_{i,n}} + F(s_{MP}, e_{\alpha_i, \gamma})^T) \cdot \langle x \rangle^1 - F(s_{MP}, e_{\alpha_i, \gamma})^T \cdot \langle x \rangle^1 = e^T_{f_{i,n}} \cdot \langle x \rangle^1$, and the $CS_2$ gets $\langle x_{i} \rangle^2 = e^T_{f_{i,n}} \cdot \langle x \rangle^2$ similarly.
We put the proof of $R_i$ calculated in the phase corresponding to the final result as follows.

\begin{footnotesize}
\begin{equation}
  \begin{split}
  R_i & = \langle R_i \rangle^1 + \langle R_i \rangle^2   \\
      & = Decrypt(mpk, \llbracket a_i, b_i \rrbracket, Sk_{\langle v_i \rangle^1}) + Decrypt(mpk, \llbracket a_i, b_i \rrbracket, Sk_{\langle v_i \rangle^2}) \\
      % &    
      & = a_i\langle 1 \rangle^1 + b_i \langle x_{i} \rangle^1 + a_i\langle 1 \rangle^2  + b_i \langle x_{i} \rangle^2 \\
      % & = a_i\langle 1 \rangle^1 + a_i\langle 1 \rangle^2 + b_i \langle x_{i} \rangle^1 + b_i \langle x_{i} \rangle^2 \\
      & = a_i(\langle 1 \rangle^1 + \langle 1 \rangle^2) + b_i(\langle x_{i} \rangle^1 + \langle x_{i} \rangle^2) \\
      & = a_i + b_i(\mathsf{Fs}_i(\langle x \rangle^1)+ \langle x_{d_i} \rangle^1 + \mathsf{Fs}_i(\langle x \rangle^2)+ \langle x_{d_i} \rangle^2) \\
      & = a_i + b_i((e^T_{f_{i,n}} + F(s_{MP}, e_{d_i, \gamma})^T) \cdot \langle x \rangle^1 - F(s_{MP}, e_{d_i, \gamma})^T \cdot \langle x \rangle^1 \\
      &  + (e^T_{f_{i,n}} + F(s_{MP}, e_{d_i, \gamma})^T) \cdot \langle x \rangle^2 - F(s_{MP}, e_{d_i, \gamma})^T \cdot \langle x \rangle^2) \\
      & = a_i + b_i(e^T_{f_{i,n}} \cdot (\langle x \rangle^1 + \langle x \rangle^2)) \\
      & = a_i + b_i(e^T_{f_{i,n}} \cdot x) \nonumber
  \end{split}
\end{equation}
\end{footnotesize}

Therefore, $R_i$ is the result of the linear function on the node $\mathbf{IN_i}$, where $(e^T_{f_{i,n}} \cdot x)$ is the feature value of the corresponding feature index on this node. Now we can compare $R_i$ with 0.5 to get the result $R_i=0$ or 1 of the corresponding boolean test function.
Hence, the protocol of the secure internal node evaluation phase is correct and calculates the direction of node $\mathbf{IN_i}$ correctly.

\subsection{Security Analysis}

In this section, we formally analyze the security of our protocol under the \textit{semi-honest adversary model} described in Section~\ref{sec:TM}. Our goal is to prove that the protocol securely realizes the ideal functionality $\mathcal{F}$, which takes as input a private decision tree model $T$ from the model provider and a private input $x$ from the data user, and returns the prediction result $f_T(x)$ to the user only.

We adopt the \textit{simulation-based security paradigm} in the \textit{stand-alone setting}. A protocol is said to be secure if, for every probabilistic polynomial-time (PPT) adversary $\mathcal{A}$ corrupting a party, there exists a PPT simulator $\mathcal{S}$ such that $\mathcal{A}$'s view in the real protocol execution is computationally indistinguishable from the simulated view generated using only the party's input and allowed outputs.

We formally analyze the security of our protocol with respect to the privacy goals defined in Section~\ref{sec:TM}. For completeness, we briefly recall them here:
\begin{itemize}
    \item \textbf{Input privacy:} The user’s input $x$ remains hidden from the model provider and the cloud servers;
    \item \textbf{Output privacy:} The prediction result $f_T(x)$ is revealed only to the user and hidden from all other parties;
    \item \textbf{Model and path privacy:} The structure and parameters of the decision tree $T$, including feature indexes, thresholds, and prediction paths, remain hidden from the user and the cloud servers.
\end{itemize}

We now present three theorems, each addressing one of the privacy properties above.

\begin{theorem}[Input Privacy]
\label{theorem1}
Assume that the pseudorandom function (PRF) $F$ is secure, the symmetric key encryption (SKE) scheme is PCPA-secure, and the functional encryption scheme (PKE-IP FE) is IND-secure. Furthermore, assume that the permutation applied to decision node indexes is generated via a Fisher-Yates shuffle seeded by outputs of $F$, such that the resulting permutation is computationally indistinguishable \footnote{We assume that the permutation of node indexes is derived via Fisher-Yates shuffle with a PRF-based seed. This is equivalent to assuming that the output permutation is a computationally indistinguishable random permutation (PRP), under the PRF security.} from uniform, i.e., it forms a secure pseudorandom permutation (PRP).
Then, for any static, semi-honest adversary $\mathcal{A}$ corrupting either $CS_1$ or $CS_2$, there exists a probabilistic polynomial-time simulator $\mathcal{S}$ such that:
\[
\mathsf{View}_\mathcal{A}^{\text{real}}(x, T) \approx_c \mathsf{Sim}_\mathcal{S}(f_T(x)),
\]
i.e., the adversary learns nothing about the user input $x$ beyond what is revealed by the prediction output $f_T(x)$.
\end{theorem}

\begin{proof}
We use a standard hybrid argument, defining a series of hybrids $\mathsf{hyb}_1$ to $\mathsf{hyb}_6$, where each hybrid gradually replaces part of the real execution with simulated values. Each step maintains computational indistinguishability under the assumed cryptographic primitives.

\begin{itemize}
    \item[\textbf{hyb1:}]\textit{PRF-seeded shuffle replacement.}  
    The simulator replaces the PRF-derived seed used to generate the Fisher-Yates shuffle with a truly uniform random seed $s'$. This results in a new node permutation $\pi'$ used to re-index internal nodes (i.e., mapping $d_i \rightarrow d_i'$).  Since the original permutation is generated via a secure PRF, its output is computationally indistinguishable from a truly random permutation. Therefore, the adversary cannot distinguish this hybrid from the real protocol.

    \item[\textbf{hyb2:}]\textit{PRF seed randomization.} 
    The PRF seed $s_{MP}$ is replaced with a random vector $s_{MP}'$. The masked feature vectors are then computed as:
    \[
    \overline{e_{f_i,n}} = e_{f_i,n} + F(s_{MP}', e_{d_i',\gamma})
    \]
    The PRF security guarantees that this step is indistinguishable to the adversary.
    
    \item[\textbf{hyb3:}]\textit{Secret-sharing simulation.} 
    The simulator replaces the user's random mask $r_i$ with a uniformly random value $r_i'$, and constructs Shamir secret shares:
    \[
    \overline{\langle x \rangle}^1 + \overline{\langle x \rangle}^2 = x
    \]
    Each share is independently random, so the server cannot infer $x$ from its view.
    
    \item[\textbf{hyb4:}]\textit{Feature masking simulation.} 
    For each $j$, the simulator computes:
    \[
    \begin{aligned}
    \overline{\langle x_j \rangle}^1 &= -F(s_{MP}', e_{j,\gamma})^T \cdot \overline{\langle x \rangle}^1, \\
    \overline{\langle x_j \rangle}^2 &= -F(s_{MP}', e_{j,\gamma})^T \cdot \overline{\langle x \rangle}^2
    \end{aligned}
    \]
    These values simulate masked features using only PRF output and secret shares.

    \item[\textbf{hyb5:}] \textit{PRF offset cancellation.}
    The simulator shows that the PRF-masked features cancel during computation:
    \[
    \begin{aligned}
    \overline{\langle x_i \rangle}^1 
    &= (e_{f_i,n}^T + F(s_{MP}', e_{d_i',\gamma})^T) \cdot \overline{\langle x \rangle}^1 \\
    &\quad - F(s_{MP}', e_{d_i',\gamma})^T \cdot \overline{\langle x \rangle}^1 \\
    &= e_{f_i,n}^T \cdot \overline{\langle x \rangle}^1
    \end{aligned}
    \]
    Similar logic applies for $\overline{\langle x_i \rangle}^2$. The masking does not leak information due to perfect cancellation.

    \item[\textbf{hyb6:}]\textit{Inner product simulation.} 
    The simulator generates simulated FE secret keys for masked vectors:
    \[
    \overline{\langle v_i \rangle}^1 = [\overline{\langle 1 \rangle}^1, \overline{\langle x_i \rangle}^1],\quad
    \overline{\langle v_i \rangle}^2 = [\overline{\langle 1 \rangle}^2, \overline{\langle x_i \rangle}^2]
    \]
    Then simulates inner product computation:
    \[
    \overline{\langle R_i \rangle}^1 = a_i \cdot \overline{\langle 1 \rangle}^1 + b_i \cdot \overline{\langle x_i \rangle}^1
    \]
    and similarly for share 2. Their sum reconstructs the true result $R_i$, preserving correctness while hiding $x$.
\end{itemize}

Since each hybrid step is computationally indistinguishable from the previous under standard assumptions (IND-CPA, PRF, Fisher-Yates pseudorandomness, and secret sharing security), the final simulated view is indistinguishable from the real execution. Thus, input privacy is preserved.
\end{proof}

\begin{theorem}[Output Privacy]
\label{theorem3}
Under the same assumptions, for any semi-honest adversary corrupting the model provider, $CS_1$, or $CS_2$, the prediction result $f_T(x)$ is only revealed to the data user. There exists a simulator $\mathcal{S}$ such that:

\[
\mathsf{View}_\mathcal{A}^{\text{real}}(x, T) \approx_c \mathsf{Sim}_\mathcal{S}(\bot),
\]

\noindent
i.e., the adversary’s view is indistinguishable from a simulation that does not know the output.
\end{theorem}

\begin{proof}[Proof sketch]
We construct a sequence of hybrid experiments to demonstrate that the adversary’s view is indistinguishable from a simulation that does not know the output.

$\mathsf{O_0.}$ The real execution where the encrypted prediction result $\textsf{Enc}(f_T(x))$ is sent to the user.

$\mathsf{O_1.} $Replace $\textsf{Enc}(f_T(x))$ with the encryption of a random value of the same length, i.e., $\textsf{Enc}(r)$ for uniformly random $r$.  
By the IND-CPA security of the symmetric encryption scheme, this hybrid is computationally indistinguishable from $\mathsf{O_0}$ for all adversaries who do not have the decryption key.

$\mathsf{O_2.}$ Replace $\textsf{Enc}(r)$ with a fixed dummy ciphertext $\textsf{Enc}(0^\lambda)$, where $\lambda$ is the security parameter. Again, by IND-CPA security, this hybrid is indistinguishable from $\mathsf{O_1}$.

In hybrid $\mathsf{O_2}$, the simulator $\mathcal{S}$ can generate all views of the adversary without knowing the actual output $f_T(x)$, as all messages are encrypted or masked.

Thus, $\textsf{View}_\mathcal{A}^{\text{real}} \approx_c \textsf{View}_\mathcal{A}^{\text{sim}}$, and output privacy is preserved.
\end{proof}

\begin{theorem}[Model and Path Privacy]
\label{theorem2}
Under the same cryptographic assumptions as Theorem~\ref{theorem1}, for any semi-honest adversary corrupting the user $DU$, or one of the cloud servers $CS_1$ or $CS_2$, there exists a simulator $\mathcal{S}$ such that:

\[
\mathsf{View}_\mathcal{A}^{\text{real}}(x, T) \approx_c \mathsf{Sim}_\mathcal{S}(f_T(x)),
\]

\noindent
i.e., the adversary learns nothing about the model structure, node parameters, or full prediction path beyond what is implied by the prediction output $f_T(x)$.
\end{theorem}

\begin{proof}[Proof sketch]
This proof builds upon Theorem~\ref{theorem1} and provides additional reasoning based on the protocol design.

\textbf{Hybrid Argument for Model Structure Privacy.}
We define a sequence of hybrids $\mathsf{H}_0 \rightarrow \mathsf{H}_1 \rightarrow \mathsf{H}_2$:
\begin{itemize}
    \item $\mathsf{H}_0$: Real execution using actual encrypted feature indexes and thresholds.
    \item $\mathsf{H}_1$: Replace encrypted indexes and thresholds with encryption of random values (same length), using IND-CPA security.
    \item $\mathsf{H}_2$: Replace with simulated ciphertexts generated by $\mathcal{S}$.
\end{itemize}
By IND-CPA security of the encryption schemes, $\mathsf{H}_0 \approx \mathsf{H}_1 \approx \mathsf{H}_2$.

\textbf{Hybrid Argument for Path Privacy.}
Define hybrid steps for each round $r$ of the decision:
\begin{itemize}
    \item $\mathsf{P}_0^r$: Real direction computation using functional encryption.
    \item $\mathsf{P}_1^r$: Replace direction with random direction.
    \item $\mathsf{P}_2^r$: Replace direction with simulator output (e.g., fixed branch).
\end{itemize}
Due to the use of masking, Fisher-Yates shuffle, and alternation between $CS_1$ and $CS_2$, no party can distinguish this replacement.
\looseness=-1

Thus, the model structure, node parameters, and prediction path remain hidden, completing the proof of simulation-based model and path privacy.
\end{proof}

Together, Theorems~\ref{theorem1},~\ref{theorem3}, and~\ref{theorem2} establish that our protocol securely realizes the ideal functionality for outsourced decision tree inference under the simulation-based security model.
\looseness=-1

\section{Performance Evaluation}

\subsection{Experimental Setup}
We implement and evaluate our protocol to demonstrate its practicality. The implementation is in Python, using the Scikit-Learn library to generate the decision tree. We use five datasets from the UCI machine learning repository \cite{asuncion2007uci} to train decision tree classifiers, i.e., Heart-disease, Credit-screening, Breast-cancer, Housing, and Spambase. All experiments are performed on an Intel Core i7 CPU @2.2GHz, ignoring network latency. 
In our experiments, we test the decision tree with realistic configurations that could arise in practice, following prior works \cite{tai2017privacy}, \cite{wu2015privately}. 
In particular, the depth $d$ ranges from 3 to 17, and the number $n$ of features ranges from 9 to 57. The parameter $l$ for the rang $\mathbb{Z}_{2^l}$ is set to 64. Note that the correctness of our design is guaranteed by the cryptographic primitives. So we focus on examining the computation and communication performance.

\subsection{Experimental Evaluation}
In this section, we evaluate the performance of our protocol. As in previous works \cite{cong2022sortinghat} \cite{tueno2019private} \cite{ma2021let} with the same $d$ and $n$, we test the computation time and communication cost for each entity. 
\looseness=-1

\begin{table}[h]
\caption{Computation performance of the model provider (in ms).}
\label{computation_mp}
\renewcommand{\arraystretch}{1.5}
\setlength{\tabcolsep}{0.8mm}{
\begin{tabular}{cccccc}
\toprule
Dataset          & Depth $d$ & Features $n$ & Internal nodes & Leaf nodes & Total    \\ \midrule
Heart-disease    & 3     & 13       & 0.039          & 0.026      & 0.065    \\
Credit-screening & 4     & 15       & 0.082          & 0.053      & 0.135    \\
Breast-cancer    & 8     & 9        & 1.268          & 0.882      & 2.150    \\
Housing          & 13    & 13       & 40.024         & 26.682     & 66.706   \\
Spambase         & 17    & 57       & 691.303        & 423.702    & 1115.005 \\ \bottomrule
\end{tabular}}
\end{table}

\subsubsection{Computation performance}
We evaluate the computational overhead of each entity.
Firstly, we evaluate the overhead of executing the protocol $\pi_{SMP}$ on $MP$. As shown in \autoref{computation_mp}, we train decision trees with various parameters on five datasets. The ``Internal nodes'' denotes the computation time on internal nodes, while ``Leaf nodes'' represents the time spent on leaf nodes. It can be observed from \autoref{computation_mp} that $MP$ takes more time to process internal nodes than leaf nodes, and this difference becomes more significant as the tree depth $d$ increases. This is because internal nodes require $MP$ to process thresholds and feature indexes, whereas leaf nodes only involve processing leaf values. Moreover, the greater the depth, the more internal and leaf nodes $MP$ need to be processed. Specifically, a tree of depth $d$ has $2^d$ leaf nodes and $2^d-1$ internal nodes, leading to exponential growth in the computational overhead. 
It is worth noting that the overhead of executing the $\pi_{SMP}$ protocol on $MP$ is independent of the number of features $n$.
Nevertheless, even for a tree with depth 17, $MP$ completes the entire protocol execution in only $1.115s$, demonstrating strong efficiency.
\looseness=-1

Secondly, we evaluate the computational overhead for $DU$ to convert data into secret shares. As described in Section \ref{input_shares}, the operations performed by $DU$ are obviously related to the number of features $n$. Moreover, the $DU$ needs to process $\gamma$ instances of $x_j$ to conceal feature indexes of internal nodes ($\gamma = 2^d -1$). \autoref{computation_du} reports the computation time required by $DU$ to process both $x$ and $x_j$. From the table, we observe that when the number of features $n$ is similar, the processing time for $x$ is comparable. However, when $n=57$, the computation time for processing $x$ is significantly higher than when $n=13$. Notably, the computation time for $x_j$ increases exponentially, as the number of $x_j$ grows exponentially with the depth $d$. For trees with $d$ below 13, the processing time of $DU$ is no more than $0.24s$. Even for a tree with $d=17$ and $n=57$, the $DU$'s runtime does not exceed $4s$, which is entirely acceptable for real-world deployments.

\begin{figure}[t]
\centering
\setlength{\abovecaptionskip}{0cm}
\centerline{\includegraphics[scale=.23]{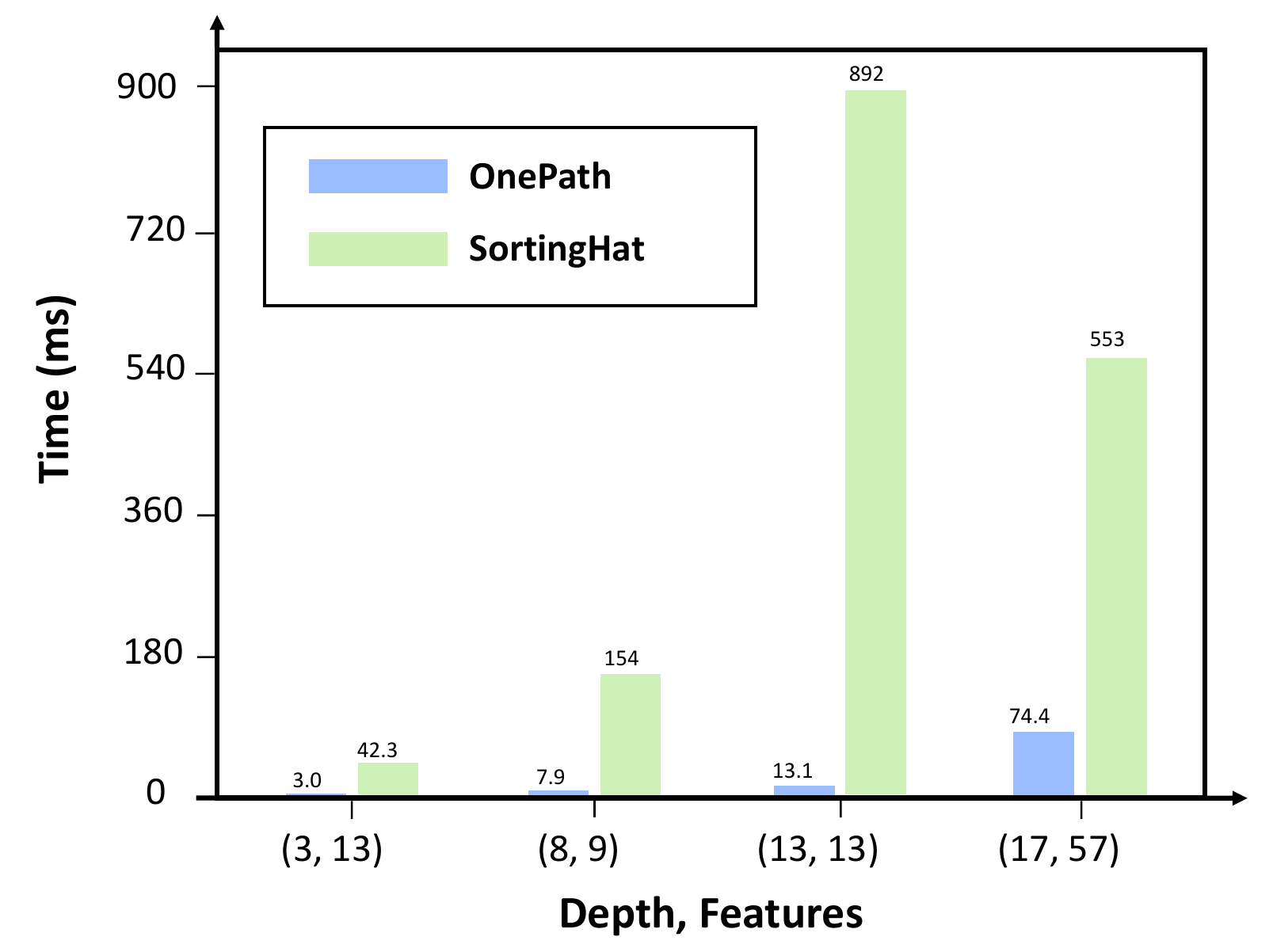}}
\caption{Computation performance of the servers (in ms).}
\label{computation_cs}
% \vspace{-2em}
\end{figure}

Thirdly, we evaluate the computational overhead on the two servers during prediction. We compare our \sysname with SortingHat \cite{cong2022sortinghat}, and the results are shown in \autoref{computation_cs}. When the number of features is similar, the time overhead grows approximately linearly since $CS_1$ and $CS_2$ only need to traverse $d$ internal nodes. In addition, during the secure feature selection phase, the servers need to cooperate with each other to compute feature values corresponding to internal nodes. As the $n$ increases, the computation time also increases. SortingHat (CCS 2022) incurs $7.4$ to $68.1$ times more overhead on the servers compared to our approach. We emphasize that a computation time of less than $75ms$ implies a very short waiting time for the $DU$, which significantly enhances the practicality of the system. Moreover, \sysname supports user-offline, requiring no user involvement during the inference phase, further improving usability.

\begin{table}[t]
\caption{Computation performance of the data user (in ms).}
\renewcommand{\arraystretch}{1.5}
\label{computation_du}
\begin{center}
\setlength{\tabcolsep}{3mm}{
\begin{tabular}{ccccc}
\hline
Depth $d$ & Features $n$ & $x$     & $x_j$     & Total    \\ \hline
3     & 13       & 0.066 & 0.202    & 0.268    \\
4     & 15       & 0.076 & 0.433    & 0.509    \\
8     & 9        & 0.046 & 7.344    & 7.390    \\
13    & 13       & 0.067 & 235.885  & 235.952  \\
17    & 57       & 0.291 & 3774.601 & 3774.892 \\ \hline
\end{tabular}}
\end{center}
% \vspace{-1em}
\end{table}

From the experimental results of the three entities, we find that when the $d$ is lower than 8, the servers have the highest computational overhead. When the $d$ is greater than 13, the servers have the least computation time, which is because the servers only need to determine $d$ internal nodes. On the other hand, both $MP$ and $DU$ need to process data related to $\gamma$, which increases exponentially. Note that we conduct the experiments on a low-performance laptop. In the real-world setting, with high-performance machine and paralleled programming, the running time can be greatly saved.

\subsubsection{Communication performance}
We also statistically measure the communication overheads between the entities. 
Firstly, the communication overhead of $MP$ primarily comes from transmitting the encrypted decision tree. Notably, $MP$ sends the entire encrypted tree to $CS_1$, while only sending the encrypted root node to $CS_2$.  The communication overhead grows exponentially with the depth $d$, and is independent of the number of features $n$. As shown in \autoref{communication}, when the depth is 3, the communication overhead of $MP$ is only 1.541 KB, whereas for a depth of 17, it increases to 25.6 MB. As a necessary step for outsourced inference, such communication overhead is entirely acceptable in modern network environments. Moreover, the servers can complete the setup by receiving the encrypted tree from $MP$ before providing prediction services, eliminating any deployment-related waiting time for the user during inference.

Secondly, we evaluate the communication overhead on the $DU$ during the outsourced inference process. The communication overhead of $DU$ consists of uploading secret shares and receiving encrypted prediction results. Due to processing $x_j$, $j=1,2, \cdots ,\gamma$, the communication overhead of $DU$ is highly affected by the depth $d$. Moreover, each feature corresponding to a feature value has to become secret sharings, so the communication overhead of $DU$ is also related to the $n$. As shown in \autoref{communication}, for the same number of features e.g., Heart-disease and Housing datasets, the communication overhead of $DU$ increases from 2.827 KB to 2.756 MB when the depth increases from 3 to 13. Overall, \autoref{communication} shows that for trees with $d \geq 13$, the computational and communication overheads of $DU$ are larger compared to other entities. This is because $DU$ performs operations concerning the $d$ and $n$, and its overhead is exponentially related to the $d$. Nevertheless, $DU$ can complete the data processing offline before inference, so the time and communication overheads are acceptable, making the approach practical in real-world scenarios.

\begin{table}[t]
\caption{Communication performance of \sysname (in KB).}
\label{communication}
\renewcommand{\arraystretch}{1.5}
\begin{center}
\setlength{\tabcolsep}{3mm}{
\begin{tabular}{ccccc}
\hline
Depth & Features & MP        & Servers   & DU        \\ \hline
3     & 13       & 1.541     & 1.124     & 2.827     \\
4     & 15       & 3.139     & 2.775     & 6.325     \\
8     & 9        & 51.827    & 49.553    & 73.187    \\
13    & 13       & 1638.292  & 1644.650  & 2821.643  \\
17    & 57       & 26214.319 & 26467.409 & 45083.392 \\ \hline
\end{tabular}}
\end{center}
% \vspace{-1em}
\end{table}

Thirdly, we evaluate the communication overhead during inference on the two servers, including the communication between $CS_1$ and $CS_2$, $CS_1$ and $KGC$, and $CS_2$ and $KGC$. As can be seen from \autoref{communication}, the communication overhead between $CS_1$ and $CS_2$ is similar to the $MP$'s results because $CS_1$ and $CS_2$ send multiple subtrees to each other. Assuming that $MP$ sends a tree of depth $d$, there are a total of $2^{d+1} - 1$ nodes. After obtaining the boolean test result of the internal node at each level, $CS_1$ or $CS_2$ will continue to send subtrees until it reaches a leaf node. Therefore, the total number of nodes sent is $2^d + 2^{d-1} + ... + 2 - d = 2^{d+1}-2-d < 2^{d+1}-1$. 
However, as shown in \autoref{communication}, when the depth $d$ is greater than or equal to 13, the communication overhead of the servers exceeds that of $MP$. This is because, during the execution of the protocol $\pi_{SMP}$, the servers send $\langle x_i \rangle^1$ or $\langle x_i \rangle^2$ to the $KGC$, and this overhead is related to the number of features $n$. 
Although the cost of sending subtrees by the servers is lower than that of $MP$, the servers face greater communication overhead than $MP$ as both $d$ and $n$ increase, due to $d$ interactions with the KGC during traversal. As a result, at depth 17, the servers’ overhead reaches 25.85 MB, while MP’s remains at only 25.60 MB. \looseness=-1

Overall, the communication overhead of $MP$ and the servers is comparable, while the overhead incurred by the $DU$ is generally higher than both. This is primarily because the $DU$ needs to additionally prepare $\gamma$ instance of $x_j$ to assist in feature selection for internal nodes. In the following, we compare our \sysname with existing approaches from both theoretical and experimental perspectives.

\begin{table}[h]
\centering
\caption{Performance comparison with prior works.}
\renewcommand{\arraystretch}{1.5}
% \vspace{1em}
\label{comparison_experiment}
\setlength{\tabcolsep}{3mm}{
\begin{tabular}{@{}ccccc@{}}
\toprule
Depth               & Features            & Scheme   & Computation & Communication \\ \midrule
\multirow{2}{*}{8}  & \multirow{2}{*}{13} & \cite{liu2020towards} & 10.21s      & 117.26KB      \\
                    &                     & \sysname  & \textbf{7.90ms}      & \textbf{49.55KB}       \\ \hline
\multirow{2}{*}{17}  & \multirow{2}{*}{57} & \cite{zheng2020securely} & 6721.59ms      & 118.00MB      \\
                    &                     & \sysname  & \textbf{74.40ms}      & \textbf{25.85MB}       \\ \hline
\multirow{2}{*}{17} & \multirow{2}{*}{57} & \cite{yuan2024efficient} & 1.49s       & \textbf{3.81MB}       \\
                    &                     & \sysname  & \textbf{74.40ms}     & 25.85MB       \\ \hline
\multirow{2}{*}{13} & \multirow{2}{*}{13} & \cite{cong2022sortinghat} & 892.00ms    & 200.00MB      \\
                    &                     & \sysname  & \textbf{13.10ms}     & \textbf{1.61MB}       \\ \bottomrule
\end{tabular}}
\end{table}

\subsection{Comparison}
Our design offers significant advantages over previous works by allowing the server to avoid traversing all nodes and enabling the client to go offline after submitting the encrypted feature vector.
As shown in \autoref{comparison_experiment}, we mainly test the running time and communication cost between the two clouds, as the $DU$ in our scheme only needs to send the query and can then remain offline while waiting for the encrypted classification result. 
Specifically, the prior design \cite{liu2020towards}, utilizing additively homomorphic encryption (AHE), suffers from slow inference, taking $10.21s$ and $117.26KB$ communication in the low-complexity $d=8$ setting due to inefficient traversal method.
In contrast, \sysname requires only $7.90ms$, while reducing communication from $117.26KB$ to $49.55KB$, achieving a computation speedup of over $1000\times$. This leap in efficiency is primarily due to our innovative traversal method, which ensures that the inference process only traverses the correct path.
The prior asymmetric secure computation (ASS)-based scheme \cite{zheng2020securely}, although its computational efficiency improved compared to \cite{liu2020towards}, its use of ASS inherently results in a substantial communication overhead of $118.00MB$.
Our \sysname in the same configuration not only reduces the communication cost to $25.85MB$, but also further accelerates the computation time from 6.72 seconds to $74.40ms$. This demonstrates our superior capability to simultaneously optimize both communication and computation overheads when handling complex models.
Furthermore, regarding the pseudo-random function and AHE scheme \cite{yuan2024efficient}, which is primarily designed for GBDT, while it achieves a low communication overhead ($3.81MB$) in specific scenarios, its computation time still reaches $1.49s$. Crucially, \cite{yuan2024efficient} fails to protect the path privacy of a single decision tree inference. \sysname not only provides the necessary path privacy protection but also achieves an approximate $20\times$ computation speedup.
\looseness=-1

Although our performance results are not directly comparable with earlier designs in non-outsourced settings, it is interesting to see how the server benefits from our secure outsourcing approach, for which we demonstrate in \autoref{comparison_experiment}. 
The prior scheme \cite{cong2022sortinghat}, which is based on fully homomorphic encryption (FHE), offers high functionality but at the cost of extreme performance overhead. When $d=13$ and $n=13$, \cite{cong2022sortinghat}'s computation time is high at $892.00ms$, coupled with a massive $200.00MB$ communication overhead.
\sysname, in the same configuration, requires only 13.10 ms for inference and successfully compresses communication to 1.606MB. This makes OnePath nearly $68\times$ faster computationally than \cite{cong2022sortinghat} and reduces communication by two orders of magnitude. This comparison highlights OnePath's lightweight and high-efficiency advantage, providing a far more feasible and practical path for decision tree inference than the powerful but costly FHE scheme.

Overall, \sysname exhibits the best overall performance across all comparisons with prior schemes. Our scheme successfully stabilizes the computation time for privacy-preserving decision tree inference in the millisecond range, a milestone previously unattainable by works based on traditional cryptographic primitives like AHE, ASS, or FHE.

\section{Discussion}
\label{discussion}
While our \sysname achieves a practical balance between efficiency and privacy in outsourced decision tree inference, there are several limitations and potential directions for improvement.
\looseness=-1

\subsection{Limitations}
First, the protocol requires a trusted $KGC$ and involves multiple rounds of interaction between parties. Although the KGC simplifies key distribution and supports the security framework under the semi-honest model, eliminating it would make the system more practical in decentralized environments. Second, there is a trade-off between computation time and communication cost. To ensure that the servers traverse only the correct path, the $DU$ must upload a number of encrypted vectors proportional to the depth $d$, which may result in significant communication overhead when $d$ is large.

\subsection{Future Work}
The current design assumes that all parties except the KGC are semi-honest. 
A promising direction is to enhance security against malicious adversaries by incorporating cryptographic techniques such as zero-knowledge proofs or verifiable computation, thereby ensuring correctness even under active attacks.

To reduce system complexity and eliminate strong trust assumptions, we will explore a privacy-preserving decision tree evaluation framework based on a single-cloud architecture in the future work. Compared to traditional two-cloud settings, which rely on the assumption of non-colluding servers, a single-cloud solution significantly simplifies deployment and reduces the risk of collusion-based attacks. This makes the single-cloud model more suitable for practical scenarios, where maintaining two independent and non-colluding cloud providers may be infeasible.
To this end, we propose leveraging a Functional-Hiding Inner Product Encryption (FHIPE) scheme to construct our solution. In this design, one party (e.g., the user or the model provider) encrypts its input using FHIPE, while the other embeds its private data into the decryption key. Due to the function-hiding property of FHIPE, the decryption key reveals no information about the embedded data or the secret key itself. The cloud server, which does not possess either party's secret input, is only responsible for computing the encrypted inner product.
While this approach effectively mitigates the risk of collusion and improves real-world deployability, it also introduces new challenges. In particular, the construction must be carefully optimized to prevent significant performance degradation due to the increased computational burden on the encryption and decryption processes.

\section{Conclusion}
In this paper, we present a practical privacy-preserving decision tree inference protocol within a two-cloud architecture. Our scheme ensures that the cloud returns an encrypted prediction result to the data user ($DU$), who only needs to send an encrypted query. Throughout the process, neither the $DU$ nor the two-cloud has access to the model provider's ($MP$) tree model. Furthermore, the evaluation involves only the internal nodes along the predicted path. Future work will focus on enhancing the efficiency of our scheme.
% In this paper, we propose a practical privacy-preserving decision tree inference protocol in a two-cloud architecture. With our decision tree inference scheme, the cloud returns an encrypted prediction result to the $DU$ who only needs to send an encrypted query. Meanwhile, the $DU$s and two-cloud know nothing about $MP$'s tree model during this process. In addition, the evaluation only requires testing the internal nodes on the predicted path. In the future, we will focus on improving the efficiency of our scheme.

\section{Acknowledgment}
This work is supported by the National Key R\&D Program of China under Grant 2022YFB3103500, the National Natural Science Foundation of China under Grant 62020106013, the Chengdu Science and Technology Program under Grant 2023-XT00-00002-GX, and the Fundamental Research Funds for Chinese Central Universities under Grant Y030232063003002.

\bibliography{bare_jrnl_new_sample4}
\bibliographystyle{IEEEtran}

% \vspace{-5in}
% \vspace{20\baselineskip}

\begin{IEEEbiography}[{\includegraphics[width=1in,height=1.25in,clip,keepaspectratio]{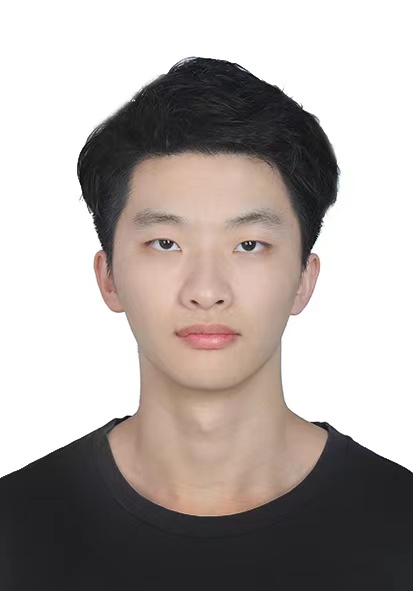}}]{Shuai Yuan}
is currently a Postdoc at the University of Electronic Science and Technology of China (UESTC) under the supervision of Prof. Binbin He. He received his Ph.D. degree from UESTC in June 2026. His research interests include applied cryptography and privacy-preserving machine learning.
\end{IEEEbiography}

\begin{IEEEbiography}[{\includegraphics[width=1in,height=1.25in,clip,keepaspectratio]{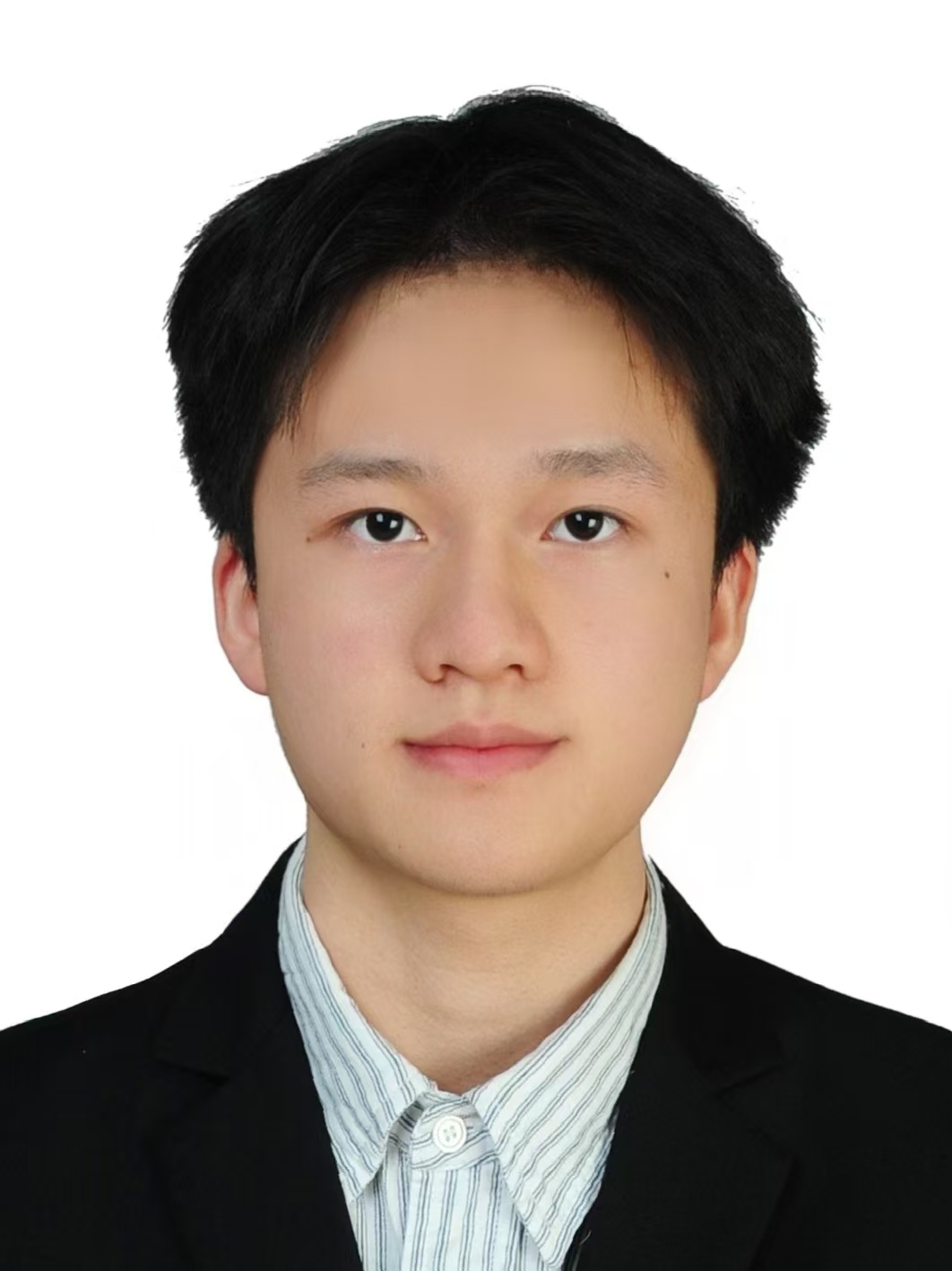}}]{Rui Zhang}
is currently working toward a Ph.D. degree with the School of Computer Science and Engineering, University of Electronic Science and Technology of China, Chengdu, China. His research interests include the intersection of AI and Machine Learning with Security and Privacy, such as machine learning security and secure multi-party computation.
\end{IEEEbiography}

\begin{IEEEbiography}[{\includegraphics[width=1in,height=1.25in,clip,keepaspectratio]{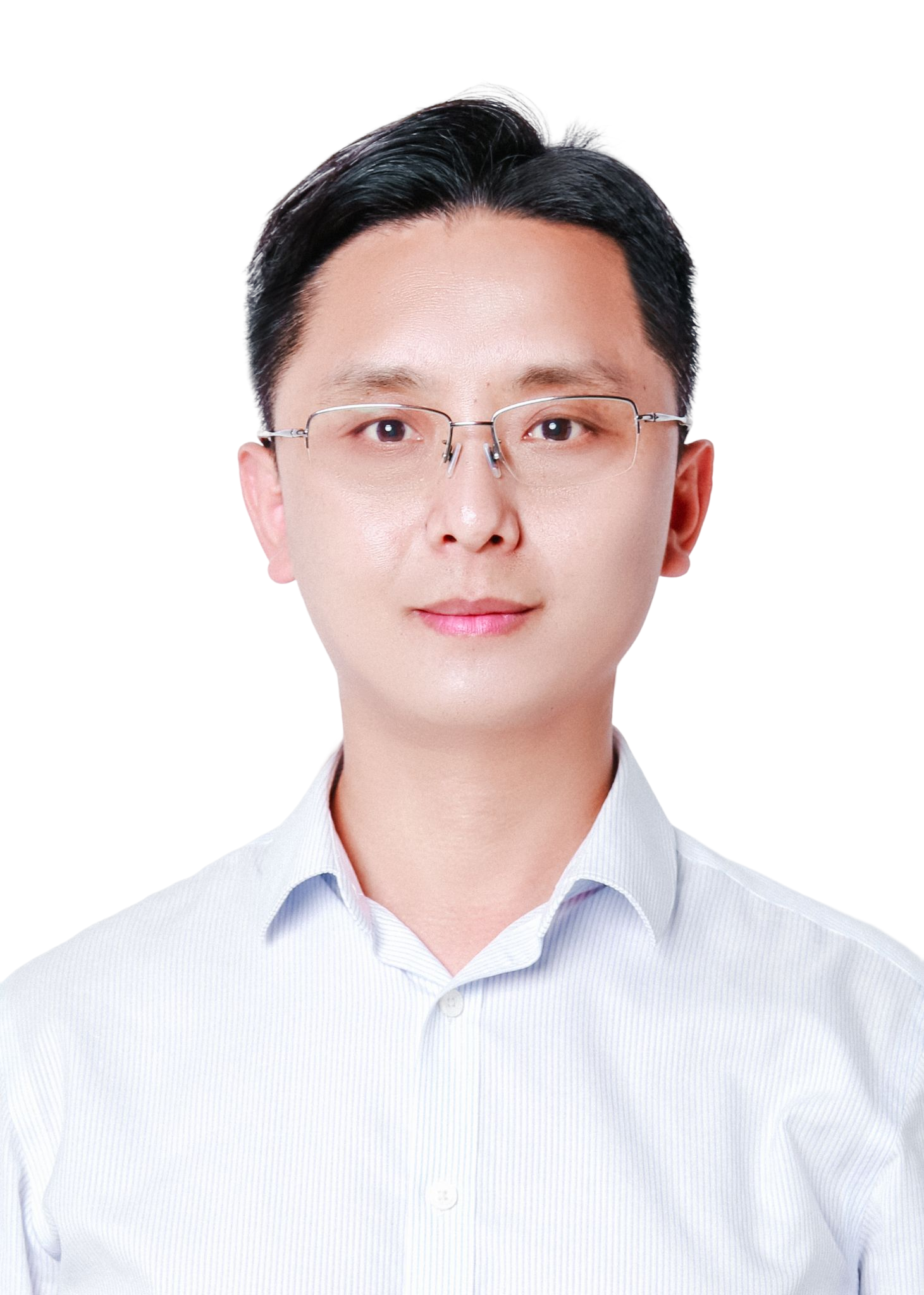}}]{Hongwei Li}
(Fellow, IEEE) is currently the Associate Dean at the School of Computer Science and Engineering, University of Electronic Science and Technology of China (UESTC). He received the Ph.D. degree from the UESTC in June 2008. He worked as a Postdoctoral Fellow at the University of Waterloo from October 2011 to October 2012. His research interests include network security and applied cryptography. Dr. Li has published more than 100 technical papers. 
He is the sole author of a book, Enabling Secure and Privacy Preserving Communications in Smart Grids (Springer, 2014). 
Dr. Li serves as the Associate Editors of IEEE Internet of Things Journal, and Peer-to-Peer Networking and Applications, the Guest Editors of IEEE Network, IEEE Internet of Things Journal and IEEE Transactions on Vehicular Technology. 
He also served the Technical Symposium Co-chairs of IEEE ICC 2022, ACM TUR-C 2019, IEEE ICCC 2016, IEEE GLOBECOM 2015 and IEEE BigDataService 2015, and Technical Program Committees for many international conferences, such as IEEE INFOCOM, IEEE ICC, IEEE GLOBECOM, IEEE WCNC, IEEE SmartGridComm, BODYNETS and IEEE DASC. 
He won Best Paper Awards from IEEE ICPADS 2020 and IEEE HEALTHCOM 2015. Dr. Li currently serves as the Vice Chair(conference) of IEEE ComSoc CIS-TC. He is the Fellow of IEEE and the Distinguished Lecturer of IEEE Vehicular Technology Society.
\end{IEEEbiography}

\begin{IEEEbiography}[{\includegraphics[width=1in,height=1.25in,clip,keepaspectratio]{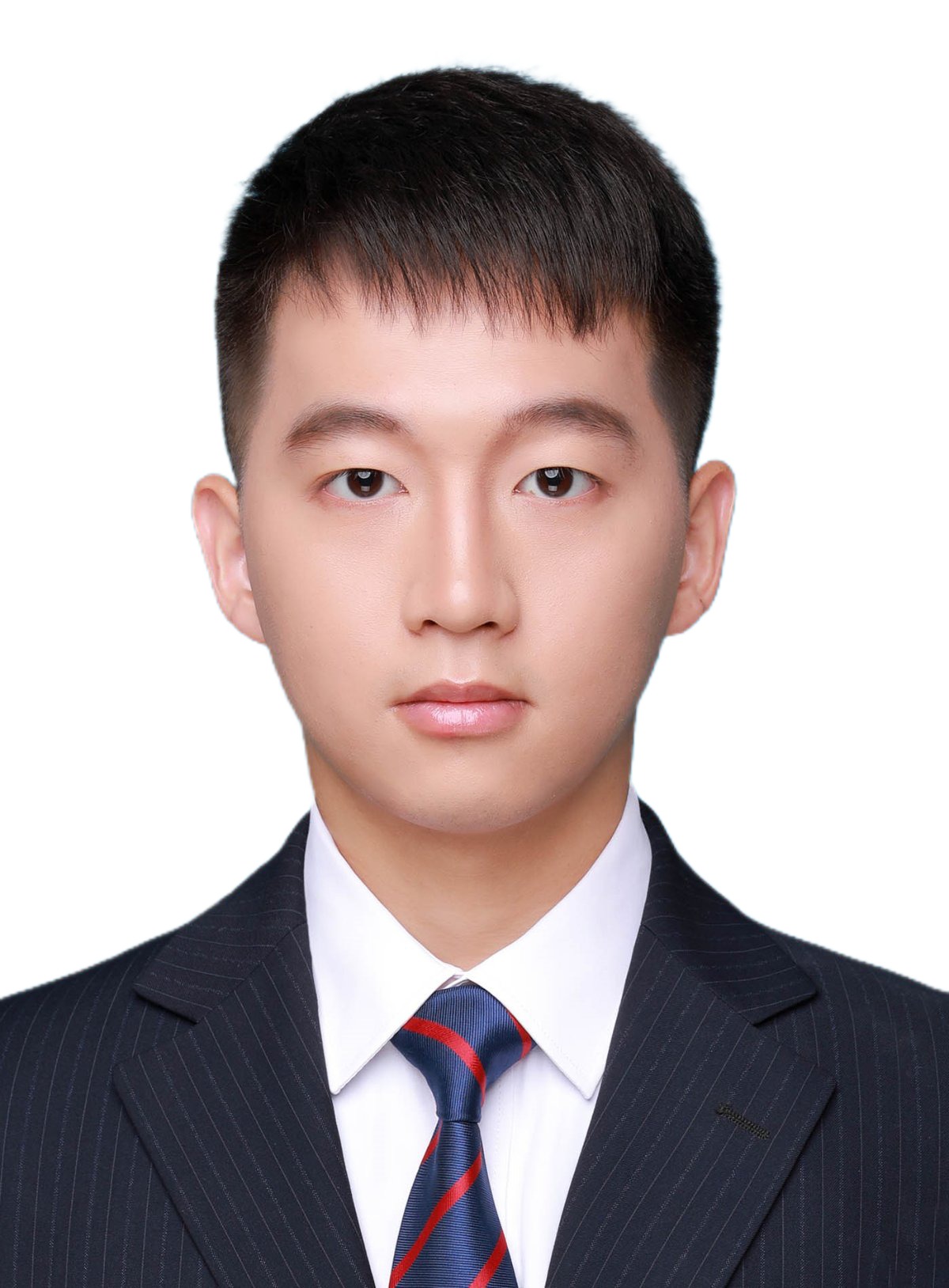}}]{Xinyuan Qian}
is currently a Postdoc at the University of Electronic Science and Technology of China (UESTC) under the supervision of Prof. Hongwei Li.
He received his Ph.D. degree from UESTC in June 2025. 
His research interests include IBE, ABE, FE, FHE, applied cryptography, and privacy-preserving machine learning.
\end{IEEEbiography}

\begin{IEEEbiography}[{\includegraphics[width=1in,height=1.25in,clip,keepaspectratio]{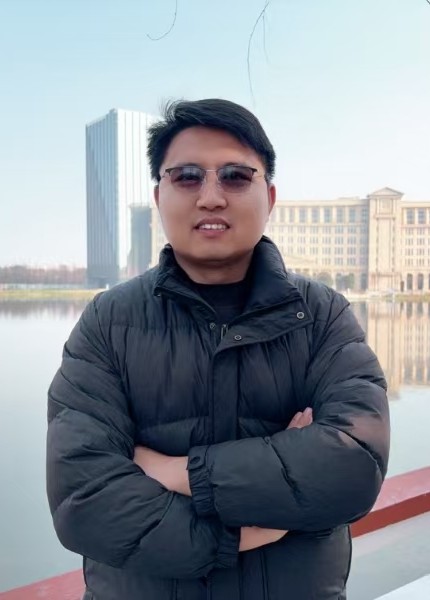}}]{Guowen Xu}
(Senior Member, IEEE) is a Full Professor at the University of Electronic Science and Technology of China (UESTC). His research focuses on cybersecurity, AI security and privacy, and applied cryptography.
He received his Ph.D. in Cyberspace Security from UESTC. Before joining UESTC, he has held positions as a Research Fellow at Nanyang Technological University, Singapore, and a Senior Research Fellow at City University of Hong Kong. 
Prof. Xu has published over 100 papers in top-tier IEEE journals and conferences, including IEEE Transactions on Dependable and Secure Computing (TDSC), IEEE Transactions on Information Forensics and Security (TIFS), IEEE INFOCOM, IEEE S\&P, ICML, and NeurIPS. 
Hiswork has been recognized with multiple awards, including the IEEE BigDataSecurity Best Paper Award (2023), IEEE ICPADS Best Paper Award (2020), Wu Wenjun First Prize of Artificial Intelligence Science and Technology Progress (2021), IEEE Early Career Speaker, IEEE Computer Society (2025), and Computing’s Top 30 Early Career Professionals, IEEE Computer Society (2025). 
Prof. Xu has extensive editorial experience, currently serving as an Associate Editor for IEEE TDSC, IEEE TIFS, IEEE/ACM Transactions on Audio, Speech, and Language Processing, IEEE Transactions on Circuits and Systems for Video Technology, and IEEE Transactions on Network and Service Management. 
He is also a Lead Guest Editor for ACM Transactions on Autonomous and Adaptive Systems (TAAS). 
Beyond his editorial roles, he has been an Area Chair or Senior Program Committee Member for prestigious international conferences, including ICML, ICLR, KDD, AAAI, NeurIPS, and CSCW.
\end{IEEEbiography}

\vfill

\end{document}